\documentclass[3p]{elsarticle}
\usepackage{hyperref}
\usepackage{ulem}
\usepackage{longtable}
\usepackage{makecell}
\usepackage{mathrsfs} 
\usepackage{amssymb,amsmath,latexsym,amsthm}
\usepackage{subfigure}
\usepackage[doublespacing]{setspace}
\usepackage{graphicx}
\usepackage{tikz,mathpazo}
\usetikzlibrary{shapes.geometric, arrows}
\usepackage{algorithm}
\usepackage{algpseudocode}
\usepackage{algorithmicx }
\usepackage{float}
\usepackage{epstopdf}
\usepackage{color}
\usepackage{dsfont}
\usepackage{multirow}
\usepackage{lineno}
\usepackage{lipsum}
\usepackage{caption}

\usepackage{multicol}
\modulolinenumbers[5]
\allowdisplaybreaks \allowdisplaybreaks[4]

\newtheorem{rmk}{Remark}[section]
\newenvironment{tequation**}{\begin{equation*}\scriptsize}{\end{equation*}}
\newenvironment{tequation***}{\begin{equation}\scriptsize}{\end{equation}}
\hypersetup{
	colorlinks=true,       
	linkcolor=black,          
	citecolor=green,        
	filecolor=magenta,      
	urlcolor=cyan,         
}

\begin{document}
	\begin{frontmatter}
		\title{A Bidomain Model for Lens Microcirculation} 
		
		\author[address1]{Yi Zhu}
		\author[address2]{Shixin Xu}
		\author[address4]{Robert S. Eisenberg}
		
		\author[address1,address2]{Huaxiong Huang \corref{mycorrespondingauthor} }
		\cortext[mycorrespondingauthor]{Corresponding author}
		\ead {hhuang@fields.utoronto.ca}

		\address[address1]{Department of Mathematics and Statistics, York University, Toronto, ON, M3J 1P3, Canada}
		
		\address[address2]{Fields Institute for Research in Mathematical Sciences, Toronto, ON, M5T 3J1, Canada}

		\address[address4]{Department of Physiology \& Biophysics, Rush University, Chicago, IL, 60612, USA}

\begin{abstract}	
 There exists a large body of research on the lens of mammalian eye over the past several decades. The objective of the current work is to provide a link between the most recent computational models  and  some of the pioneering work in the 1970s and 80s. We introduce a general non-electro-neutral model to study the microcirculation in lens of eyes. It describes the steady state relationships among ion fluxes, water flow and electric field inside cells, and in the narrow extracellular spaces between cells in the lens. Using asymptotic analysis, we derive a simplified model based on physiological data and compare our results with those in the literature. We show that our simplified model can be reduced further to the first generation models while our full model is consistent with  the most recent computational models. In addition, our simplified model captures the main features of  the full model. Our results serve as a useful link intermediate between the computational models and the first generation analytical models.  Simplified  models of this sort may be particularly helpful as the roles of similar osmotic pumps of microcirculation are examined in other tissues with narrow extracellular spaces, like cardiac and skeletal muscle, liver, kidney, epithelia in general, and the narrow extracellular spaces of the central nervous system, the ``brain".  Simplified  models may reveal the general functional plan of these systems before  full computational models become feasible and specific. 
\end{abstract}
		
\end{frontmatter}

\section{Introduction}\label{introduction}
Biological systems require continual inputs of mass and energy to stay alive. They are open systems that require flow of matter, and specific chemicals, across their boundaries. Unicellular organisms can provide that flow by diffusion to and across cell membranes. Diffusion is not adequate over distances larger than a few cell diameters, i.e., larger than say $2\times 10^{-6}$ meters, to pick a number. For that reason, multicellular organisms cannot provide those flows to their cells by diffusion itself. Multicellular organisms depend on convection to bring materials close enough to cells so diffusion to and across cell membranes can provide what the cell needs to live.

The circulatory system of blood vessels-arteries, veins, and capillaries-provides the convection in almost all tissues. But there is one clear exception, the lens of the (mammalian) eye. The lens does not have blood vessels, presumably because even capillaries would so seriously interfere with transparency. The lens is large, much larger than the length scale on which diffusion itself is efficient. The lens must provide nutrients through another kind of convection, a microcirculation of water that moves nutrients into the lens and rinses wastes out of it. The microcirculation is in fact driven by the lens itself, without an external `pump'.  The lens is itself an osmotic pump.

\begin{figure}[hpt]
	\centering
	\includegraphics[width=3.25in,height=13cm ]{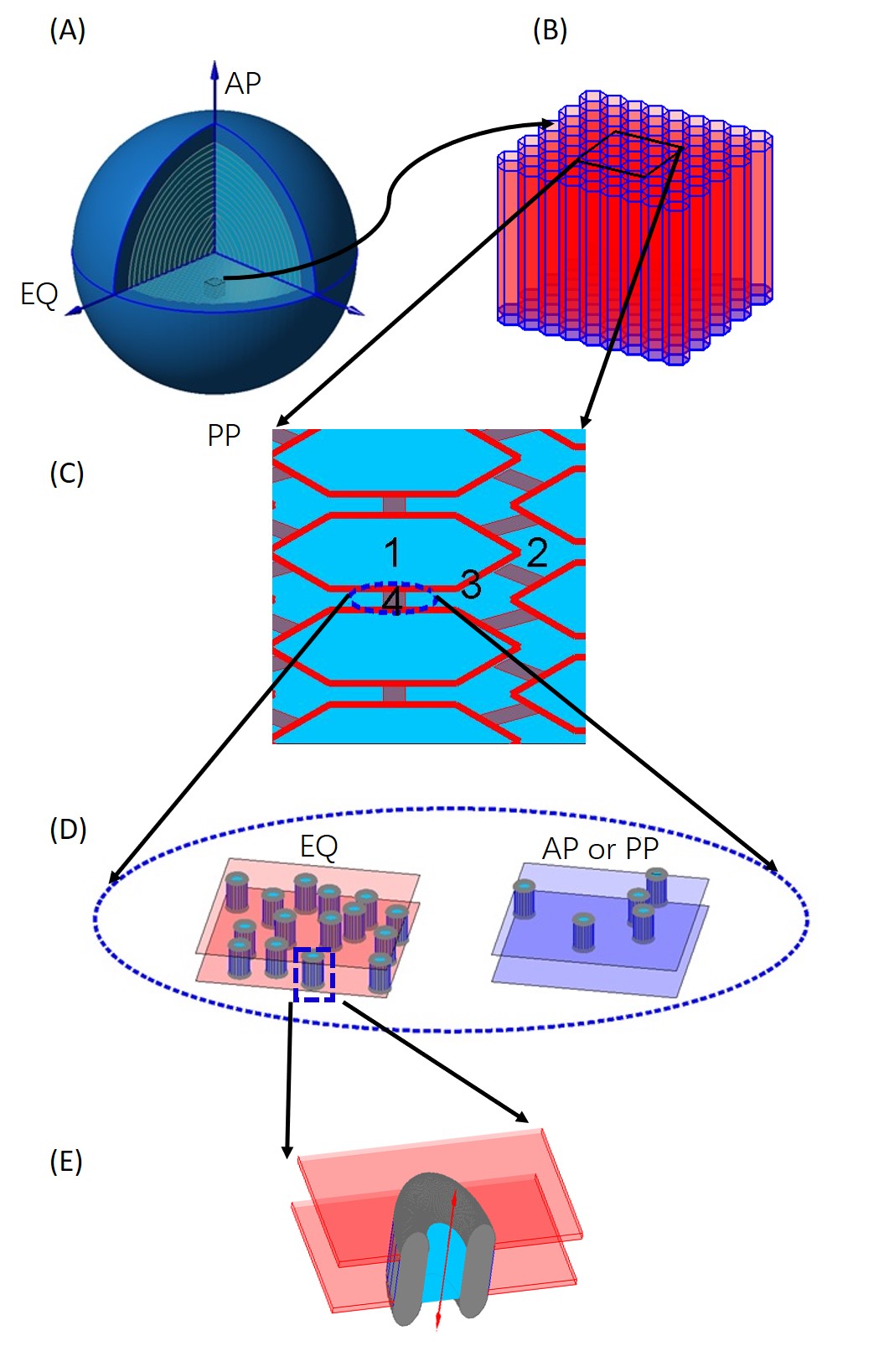}
	\caption{ Schematic diagram of  lens. (A) The sphere of the lens with three landmarks: anterior pole (AP), posterior pole (PP) and equator (EQ); (B) The control volume in the bidomain model; (C)  The micro structure of the lens: 1. intracellular region 2. extracellular region 3. cell membrane 4. gap junction (connexions);  (D) Distribution of the gap junctions between the cell membrane at EQ or AP and PP; (E) A single gap junction which allows the water and ions flows.  \label{fig:schmatic}}
\end{figure}

The lens is an asymmetrical electrical syncytium in which all cells are electrically coupled one to another, with a narrow extracellular space between the cells (see Fig.\ref{fig:schmatic}).
The extracellular space is filled with ionic solution in ‘free diffusion’ with the plasma outside cells. It may also contain specialized more or less immobile proteins and specialized polysaccharides, as well as containing obstructions formed by the connexin proteins themselves.   The intracellular space behaves very much as a large single cell would, with the bio-ions of classical electro-physiology $\mathrm{(Na^{+}, K^{+}, Cl^{-})}$ free to move without much resistance from cell to cell, and many solutes of significant size (say with diameter less than 1.5 nm) able to move as well. The intracellular media contains proteins particularly
the crystallins responsible for the high refractive index
of the lens. 
So  the lens is an example of a bidomain tissue that has been studied in some detail, first in skeletal muscle, then in cardiac muscle, and syncytia in general.  Electrical models of bidomain tissues have been developed and a general approach combining morphology, theory, and experiments has been applied in reference \cite{RN931}, showing how the lens could be studied in this tradition.

 A general approach to bidomain tissues was  implemented \cite{eisenberg2015electrical} involving detailed measurements of morphology (best done with statistical sampling by stereo-logical methods \cite{RN25704}), impedance spectroscopy \cite{RN28349, RN925, RN23185, RN926, RN929, RN28533, RN25703} using intracellular probes (micro-electrodes) that force current to flow across membranes to the extracellular baths \cite{RN23981, RN931, RN916, RN23992, RN913, RN410}, electric field theory to develop models appropriate to the structure \cite{RN575, RN928, RN927, RN921, RN919} analyzing the spectroscopic data with the field theory \cite{RN930, RN936} and checking that parameters change appropriately (i.e.,  estimates of membrane capacitance are constant) as extracellular solutions are changed in composition and concentration \cite{RN919, RN918}. This work was extended to deal with transport by Mathias and co-workers \cite{RN23951, RN23929, RN23952, RN23944, RN23924, RN23954, RN23937, RN8628, RN8504, RN8555, RN8596, RN23940, RN23931, RN23946, RN8559, RN23943, RN23955} and   computational models of the water flow in the lens were later developed in some detail \cite{RN23929, RN28404, RN28392} and exploited with great success, reviewed in \cite{ RN28406, RN28415}, also see \cite{  RN28395, RN28412, RN23921, RN28399, RN28407}. 

 The original work on electrical models is cited here because it provides coherent support, involving a range of techniques and approaches, to the general view of syncytial tissues, used here and in later work. It also shows the range of approaches needed to establish a (then) new view of a tissue.

Mathias \cite{RN23986, RN9778} realized that an asymmetrical electrical syncytium would produce convection, in particular in the lens \cite{RN8628}: he and co-workers systematically investigated the flow of water, solutes, and current in the lens, which is (in our opinion) a model of interdisciplinary research, combining theory, simulation, and measurements of many types \cite{RN23951,RN23929, RN23952, RN23944, RN23924, RN23954, RN23937,RN8504, RN8555, RN8596, RN23940, RN23931, RN23946, RN8559, RN23943, RN23955}. Computational models of the water flow in the lens were later developed in great detail \cite{RN23929, RN28404, RN28392} and compared to the more analytical models. These models have been extensively tested and we are fortunate that comprehensive reviews have been written of great value to newcomers to the field, particularly \cite{RN28406, RN28415} as well as \cite{RN28395, RN28412, RN23921, RN28399, RN28407, RN28406, RN28404, RN28415}.

Since the  pioneering  work on the models of lens microcirculation system proposed by Mathias et al. \cite{RN9778,RN930},  numerous investigations have been carried out \cite{baldo1992spatial,delamere1977comparison,RN918,RN919,RN921}. The microcirculation model has firstly relied on a combination of electrical  resistance  and current measurements and theoretical modeling \cite{RN927,mathias1985steady,RN921}. More recently, in  order  to provide a better understanding the electric current flow and potential field, the detail structure of lens has been included  \cite{malcolm2006computational,RN28404,RN28415,RN28406}, describing the asymmetric biological properties of the lens  and measurements of pressure have been made \cite{RN23921,RN23924}. Different types of  fluid  flow \cite{currie2002fundamental,ferziger2012computational} and transport properties of the ions  have  been introduced. Meanwhile,  the lens model \cite{malcolm2006computational}  has been extended to simulate age-related changes in lens physiology \cite{duncan1989human} and  a variety of physiological processes \cite{RN23952,derosa2005intercellular,shiels1998missense,mackay1999connexin46}. Reviews of  current studies on micro-circulation in lens are most helpful \cite{RN23937,RN23952,RN8555}. Despite this large body of experimental and theoretical work, it is not completely clear how they are related to each other. In particular, it is not clear how the latest computational models are related to the pioneering work, and how theoretical analysis is related to experimental findings. In this paper, we will provide such a link.

Based on the microscale model for semipermeable membrane \cite{RN28366} and bidomain method \cite{RN9778}, we construct a mathematical model  to ensure that all interactions are included  and  treated consistently. Using asymptotic analysis, we  derive a reduced model, which can be used to obtain most physiologically significant quantities except for the intracellular pressure. This simplified model  can be further  reduced to the model proposed by Mathias \cite{RN9778} with additional assumptions that  Nernst potentials (that describe gradients of chemical potential of each ionic species)   and conductance are  are constant in space. However, we will show that neither the Nernst potentials nor the conductance are constant. On the contrary, they vary significantly from the interior to the surface of the lens Therefore, both of these quantities need to be coupled as part of the solution.

 Our model also shows explicitly that the intracellular pressure is decoupled from the rest of the variables. Evolution has chosen parameters so the intracellular pressure does not affect the other parameters of the lens in a significant way. They are robust to variations of intracellular pressure. The evolutionary advantage of this adaptation is not clear to us, but may be more obvious to other workers with a greater knowledge of clinical realities that show how the lens becomes diseased \cite{gao2015feedback,RN28412,RN28399,RN28407}.
	Our simplified model suggests that all the quantities  can be computed without knowing the intracellular pressure. On the other hand, we need to solve the full model to find the value of the intracellular pressure. 
	Our model is also calibrated by experimental data and predicts the effects of gap junctions  \cite{RN23924,RN23921} described by a `membrane' permeability $\kappa_{in}$.

 Our new results extend but do not fundamentally change previous work on the lens.  We strengthen the view that the lens provides an osmotic pump to maintain the microcirculation necessary to sustain a living lens, for the life of the animal. We imagine that similar osmotic pumps create microcirculation in other cells and tissues of the body. 


%

This paper is organized as following.    The full model for micro-circulation of water and ions are proposed based on  conservation laws  in Section 2. In Section 3, we obtain the leading order model by identifying small parameters in the full model.  Based on the boundary conditions and partial differential equation (PDE) analysis, a simplified version of the leading order model is proposed and compared with the existing models.  The model calibration and simulation results are shown in Section 4. The conclusions and future work are given in  Section 5.

\section{Mathematical model} \label{Mathematical model}
\textcolor{black}{ In this section, we present a 1-D spherical symmetric non-electro-neutral model for microcirculation of  the  lens with radius $R$  by using the bidomain method  \cite{RN928}}. The model deals with two types of flow: the circulation of water (hydrodynamics) and the circulation of ions (electrodynamics), generalizing previous bidomain models that deal only with electrodynamics. \textcolor{black}{The model is mainly derived from laws of conservation of ions and water in the presence of membrane flow between intra and extracellular domains.} We note that a similar approach may be useful in other tissues with narrow extracellular space, like the heart, cardiac muscle, and the central nervous system, including the cerebral cortex, `the brain'.
\subsection{Water circulation}
We assume 
\begin{itemize}
	\item the loss of intracellular water is only through membranes flowing into the extracellular space, vice versa \cite{RN928};
	\item the trans-membrane water flux is proportional to the intra/extra-cellular hydrostatic pressure and osmotic pressure differences, i.e. \textcolor{black}{ Starling's law, classically applied to capillaries, here applied to membranes \cite{levine1967application}}. In a system like non-ideal ionic solutions in which `everything interacts with everything else' \cite{eisenberg2012life,eisenberg2013interacting}, this statement needs derivation as well as assertion. A complete, and rigorous derivation can be found in \cite{RN28366};
	\item in the rest of  this paper,  subscript $l = in,ex$  denotes the intra-/extracellular space and superscript $i =$ Na$^+$, K$^+$, Cl$^-$ denotes the $i$th specie ion.      
\end{itemize} 
\textcolor{black}{Then  we obtain the following system for intra and extracellular  velocities } in domain $\Omega = [0,R]$ 
\begin{subequations}
	\label{FS1}
	\begin{align}
	&\frac{1}{r^2}\frac{d}{dr} \left(r^2\mathcal{M}_{ex}u_{ex}\right)\nonumber\\
	&=-\mathcal{M}_{v}L_{m}\left(P_{ex}-P_{in}+\gamma_{m} k_{B}T(O_{in}-O_{ex})\right),\label{u_ex_full}\\
	&\frac{1}{r^2}\frac{d}{dr}(r^2(\mathcal{M}_{ex}u_{ex}+\mathcal{M}_{in}u_{in}))=0,  \label{fuinuex}
	\end{align}
\end{subequations}
where $u_l$ and  $P_l$   are the velocity and pressure in the intracellular and extracellular space, respectively. And $O_{l}$   is the osmotic pressure with definition 
\begin{equation*}
\label{OS1}
\begin{aligned}
&O_{ex}=\sum_{i} C^{i}_{ex},  \ \ \ \ O_{in}=\sum_{i} C^{i}_{in}+\frac{A_{in}}{V_{in}},  
\end{aligned}
\end{equation*}
where \textcolor{black}{ $C^{i}_{l}$ is the concentration of $i$th specie ion in $l$ space.}  $\frac{A_{in}}{V_{in}}$ is the density of the permanent negative charged protein. In this paper, we assume the permanent negative charged protein is uniformly distributed within intracellular space with valence of  $\bar{z}$. 
Here $\mathcal{M}_{l}$ is the ratio of intracellular area $(l=in)$ and extracellular area $(l=ex)$, $\mathcal{M}_{v}$ is the membrane area per volume unit, $\gamma_{m}$ is the intracellular membrane reflectance, \textcolor{black}{$L_{m}$ is intracellular membrane hydraulic permeability,} $k_B$ is Boltzmann constant and $T$ is temperature. 

As we mentioned before, the intracellular space is a connected space, where water can flow from cell to cell through connexin proteins \textcolor{black}{joining membranes of neighboring cells,}, and the extracellular space is narrow  with a high tortuosity.  The intracellular velocity  depends on the gradients of hydrostatic pressure  and osmotic pressure \cite{RN28366,RN28404,RN9778},  and   the extracellular velocity is determined by the gradients of hydrostatic pressure  and electric potential \cite{RN28404,wan2014self},

\begin{subequations}
	\label{VD1}
	\begin{align}
	&u_{ex}=-\frac{\kappa_{ex}}{\mu}\tau_{c}\frac{d}{dr}
	P_{ex}-k_{e}\tau_{c}\frac{d}{dr} \phi_{ex}, \\
	&u_{in}=-\frac{\kappa_{in}}{\mu}\left(\frac{d}{dr} P_{in}-\gamma_{m}k_{B}T\frac{d}{dr} O_{in} \right),
	\end{align}
\end{subequations} 
where \textcolor{black}{ $\phi_{l}$ is the electric potential in the $l$ space,}  $\tau_{c}$ is the tortuosity of extracellular region and $\mu$ is the viscosity of water, $k_{e}$ is introduced to describe  the effect of electro osmotic flow,  $\kappa_{l}$ is the permeability of intracellular region ($l=in$) and extracellular region ($l=ex$), respectively.

\textcolor{black}{Thanks to Eq. \ref{VD1}, Eq.\ref{FS1}  can be treated as equation of hydraulic pressure.
	Due to the axis symmetry condition, homogeneous Neumann boundary conditions  are used for pressure at $r=0$. }
At the surface of lens $r=R$, we set the extracellular hydrostatic pressure  to be zero and the intracellular velocity is consistent with Eq.\ref{VD1}
\textcolor{black}{
	\begin{equation}\label{fluidbd}
	\left\{
	\begin{aligned}
	&\frac{\partial P_{ex}}{\partial r} = \frac{\partial P_{in} }{\partial r}= 0, &\mbox{~at~} r=0,\\
	&P_{ex}=0,  &\mbox{~at~}r=R,\\ 
	&-\frac{\kappa_{in}}{\mu}\left(\frac{d}{dr} P_{in}-\gamma_{m}k_{B}T\frac{d}{dr} O_{in} \right)\\
	&=L_{s}\left(P_{in}-\gamma_{s}k_{B}T\left(O_{in}-O_{ex}\right)\right), &\mbox{~at~}r=R,
	\end{aligned}
	\right.
	\end{equation}
}
where  $\gamma_s$ is surface membrane reflectance and $L_s$ is surface membrane hydraulic permeability.

\subsection{Ion  circulation}
With similar assumptions, the conservation of ion concentration yields the following ion flux system

\begin{subequations}
	\label{CS1}
	\begin{align}
	&\frac{1}{r^2}\frac{d}{dr} (r^2\mathcal{M}_{ex} J^{i}_{ex})=\mathcal{M}_{v}j^{i}_{m},\\
	&\frac{1}{r^2}\frac{d}{dr}(r^2(\mathcal{M}_{ex} J^{i}_{ex}+\mathcal{M}_{in} J^{i}_{in}))=0,
	\end{align}
\end{subequations} 

The ion flux in the intracellular region $J^{i}_{in}$ and  ion flux in the extracellular region $J^{i}_{ex}$ are defined as

\begin{subequations}
	\label{FL1}
	\begin{align}
	&J_{ex}^{i}\!=\!\!C_{ex}^{i}u_{ex}\!-\!D^{i}_{ex}\tau_{c}\frac{d}{dr} C^{i}_{ex}-D^{i}_{ex}\tau_{c}\frac{z^{i}e}{k_{B}T}C^{i}_{ex}\frac{d}{dr}\phi_{ex},\\
	& J_{in}^{i}\! =\! C_{in}^{i}u_{in}-D^{i}_{in}\frac{d}{dr} C^{i}_{in}-D^{i}_{in}\frac{z^{i}e}{k_{B}T}C^{i}_{in}\frac{d}{dr} \phi_{in},
	\end{align}
\end{subequations} 
\textcolor{black}{where $D^{i}_{l}$ is the diffusion coefficient of the $i$th specie ion in the $l$ space.}
The Hodgkin-Huxley conductance formulation \cite{hodgkin1952currents,hodgkin1952components}  is  used to describe the trans-membrane flux of ions across  intracellular membrane and surface membrane
\begin{subequations}
	\label{SC1}
	\begin{align}
	&j^{i}_{m}=\frac{g^{i}}{ez^{i}}\left(\phi_{in}-\phi_{ex}-E^{i}\right),\\
	&j^{i}_{s}=\frac{G^{i}}{ez^{i}}\left(\phi_{in}-\phi_{ex}-E^{i}\right)
	\end{align}
\end{subequations} 
where $E^i =\frac{k_{B}T}{ez^{i}}\log\left(\frac{C^{i}_{ex}}{C^{i}_{in}}\right)$ is the Nernst potential (an expression of the difference of chemical potential ) of $i$th specie ion.

\begin{figure}[htp]
	\centering
	\includegraphics[width=3.25in,height=10cm ]{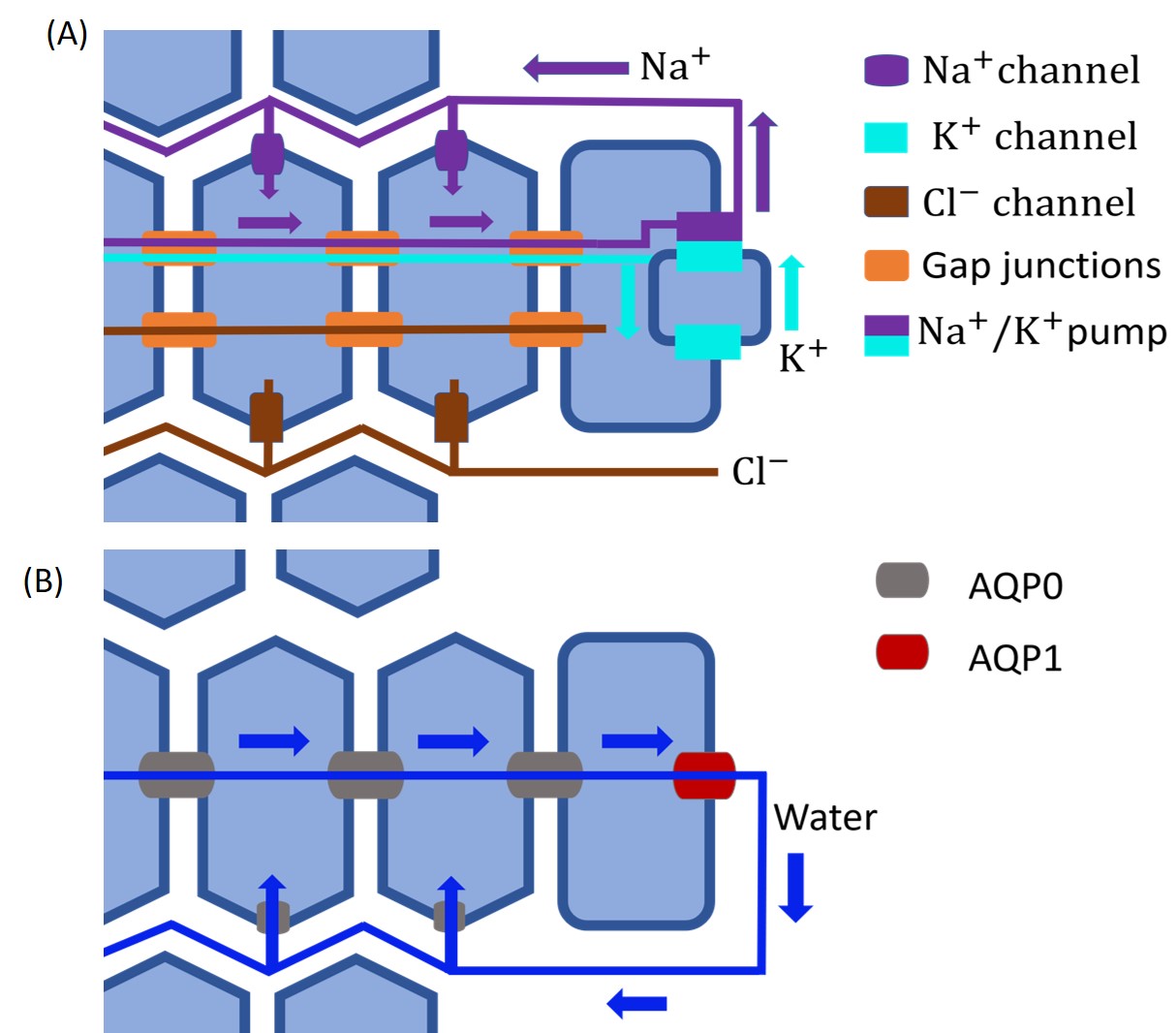}
	\caption{\textcolor{black}{(A) Schematic diagram  of ion circulation and the distributions of ion channels and pumps. The purple line represents the sodium circulation, the light green represents the potassium circulation and the brown line represents chloride circulation. The surface epithelial cells (dark black square) connect with the intracellular cells (light black hexagon) by the gap junctions (orange rectangle). The sodium and chlorine ion channels are located on the intracellular membranes, while the potassium ion channel and sodium-potassium ATP pumps are found only on the surface membrane.  (B) Schematic diagram  of water circulation. Trans-membrane water transport  is through AQP0 and AQP1 gap junctions. APQ0 gap junctions are located on the intracellular membranes, and AQP1 is on surface membrane.} \label{fig:circulation}}
\end{figure}
In \textcolor{black}{Eq. $\ref{SC1}$}, the intracellular ion conductance $g^{i}$ and surface ion  conductance $G^{i}$ depends on the ion channel distribution on the membrane (see Fig. \ref{fig:circulation}). 
Based on previous work \cite{RN9778,RN928,RN28404}, we assume that (1) only \textcolor{black}{Na$^+$ and Cl$^-$} can leak between intracellular and extracellular through ion channels inside the lens and (2) that there is no trans-membrane flux for \textcolor{black}{K$^+$} between the extracellular and intracellular region, i.e. $j^{K}_m=0$.

Similarly, homogeneous Neumann boundary conditions are used at $r=0$. At $r=R$, the extracellular concentrations are fixed and Robin boundary conditions are used for intracellular concentrations due to the trans-membrane flux and pump, 
\begin{equation}
\label{CB1}
\left\{
\begin{aligned}
&J_{ex}^i=J_{in}^i=0, &\mbox {~at~} r=0\\
&C^{i}_{ex}=C^{i}_{o},~J^{i}_{in}=j^{i}_{s}+a^{i}, &\mbox {~at~} r=R,\\
\end{aligned}
\right.
\end{equation}
where $a^{i}$ is active ion pump on the surface membrane.  Here we only consider the  sodium-potassium pump on the surface. The strength of \textcolor{black}{the pump}  depends on the ion's concentration as in \cite{RN23954,RN28404}, 
\begin{equation}
\label{PP1}
a^{Na}=3\frac{I_{p}}{e}, \ \ \  a^{K}=-2\frac{I_{p}}{e}, \ \ \
a^{Cl}=0,
\end{equation}
where 
\begin{equation}
\label{PP2}
\begin{aligned}
&I_{p}=I_{max1}\left(\frac{C^{Na}_{in}}{C^{Na}_{in}+K_{Na1}}\right)^{3}\left(\frac{C^{K}_{o}}{C^{K}_{o}+K_{K1}}\right)^{2}\\
&~~+I_{max2}\left(\frac{C^{Na}_{in}}{C^{Na}_{in}+K_{Na2}}\right)^{3}\left(\frac{C^{K}_{o}}{C^{K}_{o}+K_{K2}}\right)^{2}.
\end{aligned}
\end{equation}

Due to the  capacitance of \textcolor{black}{the} cell membrane, assumptions of exact charge neutrality can easily lead to paradoxes because they oversimplify Maxwell's equations by leaving out altogether the essential role of charge.
\textcolor{black}{We use the analysis of \cite{song2018electroneutral} and thus introduce a linear correction term replacing the charge neutrality condition \cite{RN9778,RN28404}, without introducing significant error  (See also \cite{MORI201594}),}
\begin{subequations}
	\label{NEN1}
	\begin{align}
	&(1-\eta)\left(\sum_{i}ez^{i}C_{ex}^{i}\right)=-\mathcal{M}_{v}C_{m}\left(\phi_{in}-\phi_{ex}\right),\\
	&\eta\left(\sum_{i}ez^{i}C_{in}^{i}+\bar{z}e\frac{A_{in}}{V_{in}} \right)=\mathcal{M}_{v}C_{m}\left(\phi_{in}-\phi_{ex}\right),
	\end{align}
\end{subequations} 
where $\eta$ is the porosity of intracellular region and $C_{m}$ is capacitance per unit area.

\textcolor{black}{Multiplying each ion concentration equation in Eq. \ref{CS1} with  $ez^{i}$ respectively, summing up and using Eq.\ref{NEN1}, the sodium equations are replaced by the following equations }
\begin{subequations}
	\label{PS1}
	\begin{align}
	&\frac{1}{r^2}\frac{d}{dr} \left(r^2\mathcal{M}_{ex}\left(\rho_{ex}u_{ex}-e\tau_{c}\sum_{i}D^{i}_{ex}z^{i}\frac{d}{dr} C^{i}_{ex}-\sigma_{ex}\frac{d}{dr} \phi_{ex}\right) \right)\nonumber\\
	&= \mathcal{M}_{v}\left(g^{m}\left(\phi_{in}-\phi_{ex}\right) -\sum_{i}g^{i}E^{i}\right),\\
	&\frac{1}{r^2}\frac{d}{dr} \left( r^2\mathcal{M}_{in}\left( \rho_{in}u_{in}-e\sum_{i}D^{i}_{in}z^{i}\frac{d}{dr} C^{i}_{in} -\sigma_{in}\frac{d}{dr}\phi_{in}\right) \right)\nonumber\\
	&=- \mathcal{M}_{v}\left(g^{m}\left(\phi_{in}-\phi_{ex}\right)-\sum_{i}g^{i}E^{i}\right),
	\end{align}
\end{subequations}
with boundary conditions
\begin{equation*}
\left\{
\begin{aligned}
&\frac{d\phi_{ex}}{d r}=\frac{d\phi_{in}}{d r} = 0, &\mbox{~at ~}r=0,\\ 
& \phi_{ex}=0, &\mbox{~at ~}r=R, \\
& \left(\rho_{in}u_{in} -e\sum_{i}D^{i}_{in}z^{i}\frac{d}{dr} C^{i}_{in}- \sigma_{in}\frac{d}{dr} \phi_{in}\right)\nonumber\\
&~~~ =G^{s}\phi_{in}-\sum_{i} G^{i}E^{i}+I^{\phi}_{p},&\mbox{~at ~}r=R,
\end{aligned}
\right. 
\end{equation*}
where $\rho_{in}=\frac{\mathcal{M}_{v}C_{m}}{\eta}\left(\phi_{in}-\phi_{ex}\right)+|\bar{z}|e\frac{A_{in}}{V_{in}}$ and $\rho_{ex}=\frac{\mathcal{M}_{v}C_{m}}{1-\eta}\left(\phi_{ex}-\phi_{in}\right)$
\begin{equation*}
\begin{aligned}
&g^{m}=\sum_{i}g^{i},\ \ \ \  G^{s}=\sum_{i}G^{i}, \ \ \ I^{\phi}_{p}=e\sum_{i} z^{i}a^{i}.
\end{aligned}
\end{equation*}
\textcolor{black}{In Eq. \ref{PS1}}, we define the intracellular conductance $\sigma_{in}$  and  extracellular conductance $\sigma_{ex}$ as 
\begin{equation*}
\sigma_{ex}=\frac{e^{2}\tau_{c}}{k_{B}T}\left(\sum_{i}D_{ex}^{i}(z^{i})^{2}C_{ex}^{i}\right),  \sigma_{in}=\frac{e^{2}}{k_{B}T}\left(\sum_{i}D_{in}^{i}(z^{i})^{2}C_{in}^{i}\right). 
\end{equation*}
It is obvious that system \ref{PS1} might be derived using either  Eq. \ref{CS1}  and  Eq. \ref{NEN1}. Therefore, we should drop either \textcolor{black}{Eq. \ref{CS1} or Eq. \ref{NEN1} when Eq. \ref{PS1}} is used.

\subsection{Non-dimensionalization}

Since lens circulation is driven by the sodium-potassium pump, it is natural to choose the characteristic velocity $u^*_{in}$ by the pump strength $a^{Na*}$  
\begin{equation}
\begin{aligned}
u^{*}_{in}=\frac{a^{Na*}}{O^{*}}.
\end{aligned}
\end{equation}
where $O^{*}=2\left(C^{Na}_{o}+C^{K}_{o}\right)$ is characteristic  osmotic pressure. Using Eq.\ref{fuinuex}, we obtain the scale of $u_{ex}$ as
\begin{equation}
u^{*}_{ex}=\delta_{0}^{-1} u^{*}_{in}, 
\end{equation}
where $\delta_{0}=\frac{\mathcal{M}_{ex}}{\mathcal{M}_{in}}$.
With the characteristic values for $\phi$, $P$, $C^{i}$, chosen as $\frac{k_BT}{e}$, $\frac{\mu R u^{*}_{ex}}{\kappa_{ex}\tau_{c}} $, $C^{Na}_{o}+C^{K}_{o}$, 
we obtain the dimensionless system for lens problem  as follows ( Detailed derivation is given in the appendix C.)
\begin{subequations}\label{fullmodelnd}
	\begin{align} 
	& u_{ex}=-\frac{d}{dr} P_{ex}-\delta_{1}\frac{d}{dr} \phi_{ex},\label{nduex}\\
	&\delta_{2}u_{in}=-\delta_{3}\frac{d}{dr}  P_{in}+\frac{d}{dr}  O_{in},\label{nduin}\\
	&\delta_{4}\frac{1}{r^2}\frac{d}{dr} \left( r^2 u_{in}\right)=\delta_{3} \left( P_{ex}-P_{in}\right)+\left(O_{in}-O_{ex}\right),\label{ndmassin}\\
	& u_{ex}=-u_{in},\label{nduinuex}\\
	&\sum_{i}z^{i} C^{i}_{in}+\bar{z} \frac{A_{in}}{V_{in}}=\delta_{6}\left( \phi_{in}- \phi_{ex}\right),\label{ndcnin}\\
	&\sum_{i}z^{i} C^{i}_{ex}=-\delta_{7}\left( \phi_{in}- \phi_{ex}\right),\label{ndcnex}\\
	&\frac{1}{r^2}\frac{d}{dr}  \left(r^2 J^{Cl}_{ex} \right)=\frac{ \mathcal{M}_{v}^{ex}}{z^{Cl}}\left(  \phi_{in}- \phi_{ex}-E^{Cl}\right),\label{ndclex}\\
	&\frac{1}{r^2}\frac{d}{dr}  \left(r^2 J^{Cl}_{in} \right)=-\frac{ \delta_{8}}{r^2}\frac{d}{dr}  \left(r^2 J^{Cl}_{ex}\right), \label{ndclin} \\
	&\frac{1}{r^2}\frac{d}{dr} \left(r^2 J^{K}_{ex} \right)=0,\label{ndkex}\\
	&\frac{1}{r^2}\frac{d}{dr} \left(r^2 J^{K}_{in}\right)=0,\label{ndkin}\\
	&\frac{1}{r^2}\frac{d}{dr}  \left(r^2 \left( Pe_{ex}\rho_{ex}u_{ex}-\sum_{i} D^{i}_{ex}z^{i}\frac{d}{dr} C^{i}_{ex}-\sigma_{ex}\frac{d}{dr} \phi_{ex}\right)\right) \notag \\  &=\mathcal{M}_{v}^{ex}\left(2\left(\phi_{in}-\phi_{ex}\right)-E^{Na}-E^{Cl}\right),\label{ndphiex}\\
	& \frac{1}{r^2}\frac{d}{dr} \left(r^2 \left( Pe_{in}\rho_{in} u_{in}-\sum_{i}D^{i}_{in}z^{i}\frac{d}{dr} C^{i}_{in}-\sigma_{in}\frac{d}{dr} \phi_{in}\right)\right)\notag\\  &=-\frac{\delta_8}{r^2}\frac{d}{dr}  \left(r^2 \left( Pe_{ex}\rho_{ex}u_{ex}-\sum_{i} D^{i}_{ex}z^{i}\frac{d}{dr} C^{i}_{ex}-\sigma_{ex}\frac{d}{dr} \phi_{ex}\right)\right),\label{ndphiin} 
	\end{align}
\end{subequations}	
\textcolor{black}{with homogeneous Neumann boundary conditions at $r=0$ and following boundary conditions at $r=1$}
\begin{equation}
\left\{
\begin{aligned}\label{fullmodelndbd}
& P_{ex}=0,\\
&\delta_{5}u_{in}=\delta_{3} P_{in}-\left(O_{in}-O_{ex}\right),\\
&C^{K}_{ex}=C^{K}_{o} ,~~ J^{K}_{in}=\frac{R_{s}}{z_{K}}\left(\phi_{in}-E^{K}\right)-a^{K}, \\
&C^{Cl}_{ex}=\widetilde{C}^{Na}_{o}+\widetilde{C}^{K}_{o}+\delta_{7}\left(\widetilde{\phi}_{in}-\widetilde{\phi}_{ex}\right),~~ J^{Cl}_{in}= 0,\\
&\phi_{ex}=0,\\
&Pe_{in} \rho_{in} u_{in}-\sum_{i}D^{i}_{in}z^{i}\frac{d}{dr} C^{i}_{in}-\sigma_{in}\frac{d}{dr} \phi_{in} \notag\\ &=\frac{R_{s}}{z_{K}}\left(\phi_{in}-E^{K}\right)+I^{\phi}_{p},
\end{aligned}
\right.
\end{equation}
where 
\begin{subequations}
	\begin{align}
	&\rho_{in}=\rho_0+\delta_{6} \left(\phi_{in}-\phi_{ex}\right),~~
	\rho_{0}=|\bar{z}|\frac{A_{in}}{V_{in}},\\
	&\rho_{ex}=\delta_{7} \left(\phi_{ex}-\phi_{in} \right),\\
	&\sigma_{l} =\sum_iD_{l}^i(z^i)^2C_{l}^i,\\
	& E^{i}=\frac{1}{z^{i}}\log\left( \frac{ C^{i}_{ex}}{ C^{i}_{in}}\right),\\
	& I^{\phi}_{p}=\frac{I_{p}R}{eD^{*}_{in}C^{*}}.\\
	& J^{i}_{l}=Pe_{l}   C^{i}_{l} u_{l}- D^{i}_{l} \left(\frac{d}{dr} C^{i}_{l}+z^{i} C^{i}_{l}\frac{d}{dr}  \phi_{l} \right).
	\end{align}
\end{subequations}

\section{Simplified model}\label{Simplified model}
\textcolor{black}{The full model given by system} \ref{fullmodelnd} with boundary condition \ref{fullmodelndbd}
is a  coupled nonlinear system. 
\textcolor{black}{In this section, we present a simplified version of the full model which captures the main features of the lens circulation. We first obtain the leading order model by identify the small parameters. And then by using boundary conditions and theoretical analysis, the leading order model with  is further simplified as only \textcolor{black}{one} PDE with  serial algebra equations. }

According to  those dimensionless parameters presented in the appendix B, we identify the scale of the parameters as follows
\begin{equation}
\begin{aligned}
\label{SmP}
&\{\delta_{1},\delta_{8}\} \subset{O(\epsilon)},\ \ \ \{\delta_{0}, \delta_{3}\} \subset O(\epsilon^{2}),\\
&\{\delta_{2},\delta_{4},\delta_{5},\delta_{6}, \delta_{7}\} \subset o(\epsilon^{2}).
\end{aligned}
\end{equation}
If we denote $\delta_9 = D_l^{Cl}-D_l^{K}$ and $\delta_{10} = D_l^{Cl}-D_l^{Na}$, $l=in,ex$, it yields
\begin{equation}
\label{SmP2}
\delta_{9} = O(\epsilon^2),~~ \delta_{10}=O(\epsilon).
\end{equation}  

\subsection{A priori estimation}
In this section, we provide the priori estimation of the $J^{Cl}_{in}$ as follows. By using the homogeneous Neumann boundary condition at $r=0$ and Eq.~\ref{ndphiin} yields
\begin{equation}\label{phiin2}
\begin{aligned}
&\frac{d}{dr} \phi_{in} = \frac{1}{ \sigma_{in}}\left( Pe_{in}\rho_{in} u_{in}+\delta_9\frac{d}{dr} C^{K}_{in}+\delta_{10}\frac{d}{dr} C^{Na}_{in}\right)  \\
&~ +\frac{\delta_8 }{ \sigma_{in}}    \left( Pe_{ex}\rho_{ex}u_{ex}+\delta_9\frac{d}{dr} C^{K}_{ex}+\delta_{10}\frac{d}{dr} C^{Na}_{ex}-\sigma_{ex}\frac{d}{dr} \phi_{ex}\right).
\end{aligned}
\end{equation}
From Eq.~\ref{phiin2}, since $Pe_{in}=O(\epsilon)$ and order of $\delta_{8}$, $\delta_{9}$, $\delta_{10}$ in Eqs.~\ref{SmP}-\ref{SmP2},  we obtain that 
\begin{equation}
\frac{d}{dr} \phi_{in} =O(\epsilon). 
\end{equation}
Meanwhile, from Eq.~\ref{nduin} we can have 
\begin{equation}
\label{Pri1}
\frac{d }{d r}O_{in}=O(\epsilon^{2}).
\end{equation}
and in the Eq.\ref{ndcnin}, we know 
\begin{equation}
\label{Pri2}
\frac{d}{d r} C^{Cl}_{in}= \frac{d }{d r}\left( C^{Na}_{in}+C^{K}_{in} \right)+o(\epsilon^{2}),
\end{equation}
With Eqs.~\ref{Pri1}-\ref{Pri2} and $\frac{A_{in}}{V_{in}}$ is constants, we \textcolor{black}{obtain}
\begin{equation}
\frac{d}{d r} C^{Cl}_{in}=O(\epsilon^{2}).
\end{equation}
\textcolor{black}{Furthermore, using Eq.~\ref{nduinuex} and boundary conditions for $C^{Cl}_{ex}$ in Eq.~\ref{fullmodelndbd} yieldds}
\begin{equation}
C^{Cl}_{in}=C^{Na}_{o}+C^{K}_{o}-\frac{1+|\bar{z}|}{2}\frac{A_{in}}{V_{in}}+O(\epsilon^{2}).
\end{equation}
From the experimental setting of lens \cite{RN9778,malcolm2006computational,RN28404}, we assume that 
\begin{equation}
C^{Na}_{o}+C^{K}_{o}-\frac{1+|\bar{z}|}{2}\frac{A_{in}}{V_{in}}=O(\epsilon).
\end{equation}
Therefore, 
\begin{equation}
C^{Cl}_{in}=O(\epsilon).
\end{equation}
In all, we claim that
\begin{equation}
\begin{aligned}
J^{Cl}_{in}&=Pe_{in}  C^{Cl}_{in}u_{in}-D^{Cl}_{in}\left(\frac{d}{dr} C^{Cl}_{in}+z^{Cl}C^{Cl}_{in} \frac{d}{dr} \phi_{in} \right)\\
&=O(\epsilon^{2}).
\end{aligned}
\end{equation}
By dropping the terms involving these small parameters,  the leading order of water circulation system \ref{nduex}-\ref{nduinuex} is as follows,
\begin{subequations}\label{wleadingorder}
	\begin{align}
	&u_{ex}^{0}=-\frac{d}{dr} P_{ex}^{0}-\delta_{1}\frac{d}{dr} \phi_{ex}^{0},\\
	& \frac{d}{dr} O^{0}_{in}=0,\\
	&O_{in}^{0}-O_{ex}^{0}=0,\\
	&u_{ex}^{0}=-u_{in}^{0}, 
	\end{align}
\end{subequations}
where the superscript `0' denotes the leading order approximation. From Eq.~\ref{wleadingorder}, we deduce $O^0_{ex}=O^0_{in}$ are constants, and the intracellular and extracellular flow are   counterflow.
And the total charge \textcolor{black}{in the leading order systems are  neutral} 
\begin{subequations}\label{neleadingorder}
	\begin{align}
	&\sum_{i}z^{i}C^{i,0}_{in}+\bar{z}\frac{A_{in}}{V_{in}}=0,\\
	&\sum_{i}z^{i}C^{i,0}_{ex}=0.
	\end{align}
\end{subequations}
Combining constant osmotic pressure and charge neutrality yields
\begin{subequations}
	\begin{align}
	&O_{ex}^0(r)=O_{in}^0(r)=2\left(C^{Na,0}_{ex}(1)+C^{K,0}_{ex}(1)\right),\label{conO}\\
	&\frac{d C^{Cl,0}_{in}}{d r} =\frac{d C^{Cl,0}_{ex}}{d r}=0\label{conCl},
	\end{align}
\end{subequations}
which means $C^{Cl,0}_{in}$ and $C^{Cl,0}_{ex}$ are constants and 
\begin{align}
\frac{d C^{Na,0}_{l}}{d r}=-\frac{d C^{K,0}_{l}}{d r}, \ \ \ l\in\{in,ex\}\label{NaK}.
\end{align} 
And the leading order of potassium and chloride   concentrations satisfy
\begin{subequations}\label{kleadingorder}
	\begin{align}
	&\frac{1}{r^2}\frac{d}{dr} \left(r^2 J^{K,0}_{in}\right)=0,\\
	&  \frac{1}{r^2}\frac{d}{dr} \left(r^2  J^{K,0}_{ex}\right)=0\\
	&  \frac{1}{r^2}\frac{d}{dr} \left(r^2 J^{Cl,0}_{ex}\right)=\frac{ \mathcal{M}_{v}^{ex}}{z^{Cl}}\left(  \phi_{in}^0- \phi_{ex}^0-E^{Cl,0}\right)\\
	& \frac{1}{r^2}\frac{d}{dr} \left(r^2 J_{in}^{Cl,0}\right)= - \frac{1}{r^2}\frac{d}{dr} \left(r^2\delta_{8} J^{Cl,0}_{ex} \right), 
	\end{align} 
\end{subequations}
where $J_{l}^{i,0}= Pe_{l}  C^{i,0}_{l}u^0_{l}-D^{i}_{l}\left(\frac{d}{dr} C^{i,0}_{l}+z^{i}C^{i,0}_{l} \frac{d}{dr} \phi_{l}^{0} \right)$ with $i = K, Cl $ and $l= in, ex,$ 
$E^{Cl,0}=\frac{1}{z^{Cl}}\log\left( \frac{C^{Cl,0}_{ex}}{C^{Cl,0}_{in}}\right).$ 

For the electric potential, using the homogeneous Neumann boundary condition at $r=0$ and Eqs. \ref{conO} -\ref{NaK}, \ref{ndphiin} yields
\begin{eqnarray}\label{phiin}
&\frac{d}{dr} \phi_{in} = \frac{1}{ \sigma_{in}}\left( Pe_{in}\rho_{in} u_{in}+\delta_9\frac{d}{dr} C^{K}_{in}+\delta_{10}\frac{d}{dr} C^{Na}_{in}\right)\\  &+\frac{\delta_8 }{ \sigma_{in}}    \left( Pe_{ex}\rho_{ex}u_{ex}+\delta_9\frac{d}{dr} C^{K}_{ex}+\delta_{10}\frac{d}{dr} C^{Na}_{ex}-\sigma_{ex}\frac{d}{dr} \phi_{ex}\right)\nonumber.
\end{eqnarray}

At the same time, based on the intracellular  equation  of potassium   Eq.~\ref{ndkin}
the homogeneous Neumann boundary condition at $r=0$ and Eqs.~\ref{conO} -\ref{NaK}, we have 
\begin{eqnarray}\label{kin}
D^{K}_{in}\frac{d}{dr}  C^{K}_{in}=\left(Pe_{in}   C^{K}_{in}u_{in}- D^{K}_{in}z^{K} C^{K}_{in} \frac{d}{dr}  \phi_{in}\right) 
\end{eqnarray}  

Substituting Eq.~\ref{phiin} into Eq.~\ref{kin} yields
\begin{eqnarray}
\label{e17}
&&\left(1-z^{K} C^{K}_{in}\frac{  \delta_{10} }{\sigma_{in}}\right)D^{K}_{in}\frac{d C^{K}_{in}}{d r}\notag\\
&=&\!\!\left(\!\!\left(1\!\!-z^{K} D^{K}_{in}\frac{ \rho_{in}}{\sigma_{in}}\right)Pe_{in}  u_{in}+z^{K} D^{K}_{in} \frac{\delta_{8}\sigma_{ex}}{\sigma_{in}}\frac{d \phi_{ex}}{d r}\right)C^{K}_{in}\notag\\
&&+O(\epsilon^2), 
\end{eqnarray}
where we used that fact that $\rho_{ex} = o(\epsilon^2)$, $\delta_{9}=O(\epsilon^2)$ and $\frac{d C_l^{K}}{d r} =-\frac{d C_l^{Na}}{d r}+O(\epsilon^2), ~~l\in\{in,~ex\} .$
Since
$Pe_{in}= O(\epsilon)$ , and $ \delta_{8} = O(\epsilon)$, in \textcolor{black}{Eq.~$\ref{e17}$}, we claim 
\begin{equation}
\label{gradNa}
\frac{d C^{K}_{in}}{d r}= O(\epsilon).
\end{equation}

Combining Eqs. \ref{phiin} and  \ref{gradNa} yields the  leading order approximation of intracellular potential 
\begin{equation}\label{phiinleading}
\frac{d}{dr} \phi_{in}^0 = \frac{1}{ \sigma_{in}^0}  Pe_{in}\rho_{0} u_{in}^0-\frac{\delta_8 }{ \sigma_{in}^0} \sigma_{ex}^0\frac{d}{dr} \phi^0_{ex}=O(\epsilon),
\end{equation}
where  $\sigma_{in}^0 =\sum_iD_{in}^i(z^i)^2C_{in}^{i,0}$, $\sigma_{ex}^0 =\sum_iD_{ex}^i(z^i)^2C_{ex}^{i,0}.$

Similarly, the leading order approximation of extracellular potential is 
\begin{equation}
\resizebox{.4 \textwidth}{!} 
{$\begin{aligned}
	&-\frac{1}{r^2}\frac{d}{dr} \left(r^2 \left(\delta_{10}\frac{d}{dr} C^{Na,0}_{ex}+\sigma_{ex}^{0}\frac{d}{dr} \phi_{ex}^0\right)\right)\\
	&~~=\mathcal{M}_{v}^{ex}\left(2\left(\phi_{in}^0-\phi_{ex}^0\right)-E^{Na,0}-E^{Cl,0}\right),
	\end{aligned}$}
\end{equation}  
where 
$E^{Na,0}=\frac{1}{z^{Na}}\log\left( \frac{ C^{Na,0}_{ex}}{ C^{Na,0}_{in}}\right) $. 

To summarize, the leading order approximation of system \ref{fullmodelnd}-\ref{fullmodelndbd} is given by, in domain $\Omega=[0,1]$ 

\begin{subequations}\label{leadingorder}
	\begin{align}
	&u_{ex}^{0}=-\frac{d}{dr} P_{ex}^{0}-\delta_{1}\frac{d}{dr} \phi_{ex}^{0},\label{luex}\\
	& \frac{d}{dr} O^{0}_{in}=0,\label{loin}\\
	&O_{in}^{0}-O_{ex}^{0}=0,\label{loinoex}\\
	&u_{ex}^{0}=-u_{in}^{0},\label{luiuuex} \\
	&\sum_{i}z^{i}C^{i,0}_{in}+\bar{z}\frac{A_{in}}{V_{in}}=0,\label{lcnin}\\
	&\sum_{i}z^{i}C^{i,0}_{ex}=0,\label{lcnex}\\
	&\frac{1}{r^2}\frac{d}{dr} \left(r^2 J^{K,0}_{in}\right)=0,\label{lkin}\\
	&  \frac{1}{r^2}\frac{d}{dr} \left(r^2 J^{K,0}_{ex}\right)=0,\label{lkex}\\
	&  \frac{1}{r^2}\frac{d}{dr} \left(r^2  J^{Cl,0}_{ex}\right)=\frac{ \mathcal{M}_{v}^{ex}}{z^{Cl}}\left(  \phi_{in}^0- \phi_{ex}^0-E^{Cl,0}\right),\label{lclex}\\
	& \frac{1}{r^2}\frac{d}{dr} \left(r^2  J^{Cl,0}_{in}\right)= - \frac{1}{r^2}\frac{d}{dr} \left(r^2\delta_{8} J^{Cl,0}_{ex} \right),\label{lclin}\\
	&\frac{d}{dr} \phi_{in}^0 = \frac{1}{ \sigma_{in}^0}  Pe_{in}\rho_{0} u_{in}^0-\frac{\delta_8 }{ \sigma^0_{in}} \sigma_{ex}^0\frac{d}{dr} \phi^0_{ex},\label{lphiin}\\
	& -\frac{1}{r^2}\frac{d}{dr} \left(r^2 \left(\delta_{10}\frac{d}{dr} C^{Na,0}_{ex}+\sigma^0_{ex}\frac{d}{dr} \phi_{ex}^0\right)\right)\notag\\
	&~~=\mathcal{M}_{v}^{ex}\left(2\left(\phi_{in}^0-\phi_{ex}^0\right)-E^{Na,0}-E^{Cl,0}\right),\label{lphiex}
	\end{align}
\end{subequations}  
with boundary conditions at $r=1$
\begin{equation}\label{leadingorderbd}
\left\{
\begin{aligned}
&P^0_{ex}=0,  C^{Cl,0}_{ex}= C^{Na}_{o}+ C^{K}_{o}, C^{K,0}_{ex}=C^{K}_{o},\\
& Pe_{in}  C^{K,0}_{in}u^0_{in}-D^{K}_{in}\left(\frac{d}{dr} C^{K,0}_{in}+z^{K}C^{K,0}_{in}\frac{d}{dr} \phi^0_{in} \right)\\ &=\frac{R_{s}}{z_{K}}\left(\phi^0_{in}-E^{K,0}\right)-a^{K},\\
& Pe_{in} \rho_{0} u_{in}^0+\delta_{10}\frac{d}{dr} C^{Na,0}_{in}-\sigma_{in}\frac{d}{dr}\phi^0_{in}\\ &=\frac{R_{s}}{z_{K}}\left(\phi_{in}^0-E^{K,0}\right)+I^{e}_{p},\\
&\phi_{ex}^0=0.
\end{aligned}
\right. 
\end{equation}
In the following, we will further simplify  Eqs. \ref{leadingorder}-\ref{leadingorderbd} and obtain the relationships between $\phi_{ex}^0$ and other leading order variables by using assumptions concerning the boundary conditions.
\subsection{Relation between $\phi^0_{in}$ and $\phi^0_{ex}$}
Combining Eqs. \ref{luex}, \ref{luiuuex}  and \ref{lphiin}, and integrating with respect to $r$ yields
the relation between $\phi^0_{in}$ and $\phi^0_{ex}$ as
\begin{equation}
\begin{aligned}
\label{r1}
&\phi_{in}^0(r)=\left(\frac{Pe_{in}\rho_{0} \delta_{1}}{\sigma^0_{in}}-\frac{\delta_{8}\sigma^0_{ex}}{\sigma^0_{in}}\right)\phi^0_{ex}(r)\\
&+\frac{Pe_{in}\rho_{0} }{\sigma_{in}^0}P^0_{ex}(r)+\phi^0_{in}(1).
\end{aligned} 
\end{equation}
where we used the  boundary conditions
$\phi^0_{ex}(1)=P_{ex}^0(1)=0.$
\subsection{Relation between $P^0_{ex}$ and $\phi^0_{ex}$}
By the homogeneous Neumann boundary condition on $r=0$ and  Eq. \ref{lclin}, we have 
\begin{equation}
\label{e11}
\begin{aligned}
&J^{Cl,0}_{in} +\delta_{8}J^{Cl,0}_{ex}=0.
\end{aligned}
\end{equation}
By \textcolor{black}{Eq.~\ref{conCl}}, we can \textcolor{black}{ divide \textcolor{black}{Eq.~\ref{e11}} by $C^{Cl,0}_{ex}$ on both sides}, we get
\begin{equation}
\label{e13}
\resizebox{.3 \textwidth}{!} 
{$ \begin{aligned}
	&\left(Pe_{in}  \frac{C^{Cl,0}_{in}}{C^{Cl,0}_{ex}}u_{in}^0-D^{Cl}_{in}z^{Cl}\frac{C^{Cl,0}_{in}}{C^{Cl,0}_{ex}} \frac{d \phi^0_{in}}{d r}  \right)\\
	&~+\delta_{8} \left(Pe_{ex} u_{ex}^0-D^{Cl}_{ex}z^{Cl}\frac{d \phi^0_{ex}}{d r} \right)=0.\end{aligned}$}
\end{equation}

Base on the charge neutrality Eq.~\ref{neleadingorder},  constant osmotic pressure Eq.~\ref{conO} and parameters in Appendix B, we denotes  
\begin{equation}
\delta_{11} = \frac{C_{in}^{Cl,0}}{C_{ex}^{Cl,0}} =\frac{C^{Na}_{o}+C^{K}_{o}-\frac{1+|\bar{z}|}{2}\frac{A_{in}}{V_{in}}}{C_{o}^{Cl,0}} = O(\epsilon).
\end{equation}

Then combining the Eqs.~\ref{phiinleading} and $Pe_{in}= O(\epsilon)$, Eq.~\ref{e13} yields  the following equation by omitting the higher order terms 
\begin{equation}
\label{RphiP}
Pe_{ex}u^0_{ex}-D^{Cl}_{ex}z^{Cl}\frac{d \phi^0_{ex}}{d r}=0.
\end{equation}
Finally, by using the boundary condition, 
we have the relation between extracellular pressure and electric potential as 
\begin{equation}
\label{r2}
P^0_{ex}=\frac{D^{Cl}_{ex}-Pe_{ex}\delta_{1}}{Pe_{ex}}\phi^0_{ex}.
\end{equation}

\subsection{Expression of $E^{Na}$}
Based on potassium equation and relation in \textcolor{black}{Eqs.~\ref{r1} and \ref{r2}},  we have  expression for $C^{K}_{in}$ and $C^{K}_{ex}$ as
\begin{subequations}
	\label{Ksolu}
	\begin{align}
	C^{K,0}_{ex}&=C^{K,0}_{0}\exp\left(-\left(1+\frac{D^{Cl}_{ex}}{D^{K}_{ex}}\right)\phi^0_{ex}\right),\\
	C^{K,0}_{in}&=C^{K,0}_{in}(1)\exp\left(\left(\frac{Pe_{in}D^{Cl}_{ex}}{Pe_{ex}D^{K}_{in}}  -\frac{Pe_{in}D^{Cl}_{ex}\rho_{0}}{Pe_{ex}\sigma^0_{in}}\right)\phi^0_{ex}\right)\notag\\
	&~~\exp\left(\left( \frac{\delta_{9}\sigma^0_{ex}}{\sigma^0_{in}}\right)\phi^0_{ex}\right),
	\end{align}
\end{subequations} 
where
\begin{equation}
C^{K,0}_{in}(1)=C^{K,0}_{o}\exp\left(\frac{a^{K}}{R_{s}}-\phi_{in}(1)\right).
\end{equation}
Based on Eq.~\ref{neleadingorder}, 
we can get 
\begin{eqnarray}
\label{Ena}
E^{Na,0}&=&\frac{1}{z^{Na}}\log\left( \frac{C^{Na,0}_{ex}}{C^{Na,0}_{in}}\right)\\
&=&\frac{1}{z^{Na}}\log \left(\frac{C^{Cl,0}_{ex}-C^{K,0}_{ex}}{C^{Cl,0}_{in}+|\bar{z}| \frac{A_{in}}{V_{in}}-C^{K,0}_{in}}\right). \nonumber
\end{eqnarray}

\subsection{Extracellular electric potential system}
By \textcolor{black}{Eqs.~\ref{r1} and \ref{r2}}, we have $\phi_{in}$ as
\begin{equation}
\label{phi_inex}
\begin{aligned}
\phi^0_{in}(r)
\!\!=\!\!\left(\frac{D^{Cl}_{ex}Pe_{in}\rho_{0} }{\sigma^0_{in}Pe_{ex}}-\frac{\delta_{9}\sigma^0_{ex}}{\sigma^0_{in}}\right)\phi^0_{ex}(r)+\phi^0_{in}(1),
\end{aligned} 
\end{equation}
The value $\phi_{in}^0(1)$ is determined by  the boundary condition of $\phi^0_{in}$ in Eq.~\ref{leadingorderbd}, where
\begin{equation}
\label{phiB} 
-\mathcal{M}_{v}^{in}\!\!\int_{0}^{1}\!\! \left(2\left(\phi^0_{in}\!\!-\!\!\phi^0_{ex}\right)\!\!-\!\!E^{Na,0}-E^{Cl,0}\right) s^{2}ds=a^{Na}.
\end{equation}
where we use 
\begin{equation*}
a^{Na}=-a^{K}+I^{\phi}_{p}, \ \ \ \ \ \frac{R_{s}}{z^{K}}\left(\phi_{in}-E^{K}\right)=-a^{K}.
\end{equation*}

To summarize, we obtained the simplified model of system \ref{leadingorder}-\ref{leadingorderbd} as follows  
\begin{subequations}
	\label{smodel}
	\begin{align}
	&-\frac{1}{r^2}\frac{d}{dr} \left(r^2 \left(\delta_{10}\frac{d}{dr} C^{Na,0}_{ex}+\sigma_{ex}^{0}\frac{d}{dr} \phi_{ex}^0\right)\right)\notag\\
	&~=\mathcal{M}_{v}^{ex}\left(2\left(\phi_{in}^0-\phi_{ex}^0\right)-E^{Na,0}-E^{Cl,0}\right),\label{sphiex}\\
	& \phi^0_{in}(r)
	=\left(\frac{D^{Cl}_{ex}Pe_{in}\rho_{0} }{\sigma^0_{in}Pe_{ex}}-\frac{\delta_{9}\sigma^0_{ex}}{\sigma^0_{in}}\right)\phi^0_{ex}(r)+\phi^0_{in}(1),\label{sphiin}\\
	&-\mathcal{M}_{v}^{in}\int_{0}^{1} \left(2\left(\phi^0_{in}-\phi^0_{ex}\right)-E^{Na,0}-E^{Cl,0}\right) s^{2}ds,\notag\\
	&=a^{Na}\label{sphiin1}\\
	&u_{ex}^{0}=-\frac{d}{dr} P_{ex}^{0}-\delta_{1}\frac{d}{dr} \phi_{ex}^{0},\label{suex}\\
	&u_{in}^{0} = -u_{ex}^{0}\label{suin}\\
	&C^{K,0}_{ex}=C^{K,0}_{0}\exp\left(-\left(1+\frac{D^{Cl}_{ex}}{D^{K}_{ex}}\right)\phi^0_{ex}\right),\label{skex}\\
	&C^{K,0}_{in}=C^{K,0}_{in}(1)\exp\left(\left(\frac{Pe_{in}D^{Cl}_{ex}}{Pe_{ex}D^{K}_{in}}  -\frac{Pe_{in}D^{Cl}_{ex}\rho_{0}}{Pe_{ex}\sigma^0_{in}}\right)\phi^0_{ex}\right)\notag\\
	&~~\exp\left(\left( \frac{\delta_{9}\sigma^0_{ex}}{\sigma^0_{in}}\right)\phi^0_{ex}\right),\label{skin}\\
	&C^{Na,0}_{ex}=C^{Cl,0}_{ex}-C^{K,0}_{ex},\label{snaex} \\ 
	&C^{Na,0}_{in}=C^{Cl,0}_{in}+\bar{z} \frac{A_{in}}{V_{in}}-C^{K,0}_{in},\label{snain}\\
	&C^{Cl,0}_{in}=C^{Na,0}_{o}+C^{K,0}_{o}-\frac{1+|\bar{z}|}{2}\frac{A_{in}}{V_{in}},\label{sclin}\\
	&C^{Cl,0}_{ex}=C^{Na,0}_{o}+C^{K,0}_{o},\label{sclex}\\
	&P_{ex}^0=\frac{D^{Cl}_{ex}-Pe_{ex}\delta_{1}}{Pe_{ex}}\phi^0_{ex}.\label{spex}
	\end{align}
\end{subequations}	
\textcolor{black}{
	with boundary conditions
	\begin{equation}
	\left\{
	\begin{aligned}
	&\frac{d \phi^{0}_{ex}}{dr}=0, &\mbox{~at ~} r=0,\\
	&\phi^{0}_{ex}=0, &\mbox{~at ~} r=1.\\
	\end{aligned}
	\right.
	\end{equation}
}
\begin{rmk}
	Under the same assumptions in \cite{RN9778}, for example, uniform diffusion constants for all ions, constant Nernst potential, our simplified model \textcolor{black}{system~\ref{smodel}}  recovers the model proposed by Mathias.  The main contribution here is that we remove the assumptions that Nernst potentials and effective conductance should be constants.  By using the relationships between ions concentrations and external potential, we obtain the space dependent Nernst potential which yields a much better approximation to the full model (see Fig.~\ref{figure4}). 
\end{rmk}

\section{Results and discussion}\label{Numerical comparison}	
In this section, we present  numerical simulations  using both the full  and simplified models. 
\textcolor{black}{ Finite Volume Method   \cite{song2018electroneutral} is used in order to preserve mass conservation of ions. The  convex
	iteration \cite{lee2010new} is employed to solve the nonlinear coupled system. The numerical algorithm is implemented in Matlab.  }

\subsection{Model calibration: membrane conductance effects intracellular hydrostatic pressure}
In this section, we first calibrate the full model by the comparing with the experimental data to study effect of connexin to intracellular  hydrostatic pressure. 

Intracellular hydrostatic pressure is an important
physiological quantity \cite{gao2015feedback}. In the paper \cite{RN23924,RN23921}, the authors showed the connexin (gap junction) conductance play an important role in the microcirculation of lens.  It is said that if the intracellular conductance $\frac{\kappa_{in}}{\mu_{in}}$ in lenses is approximately doubled, the hydrostatic pressure gradient in the lenses should become approximately half of the original one. In this section, we calibrate our model. We choose a value of the  intracellular conductance ($\frac{\kappa_{in}}{\mu_{in}}$) that correctly calculates the experimental \textcolor{black}{results} in the  \cite{RN23924,RN23921}.

\begin{figure*}[htp]
	\centering
	\includegraphics[width=6in,height=10cm ]{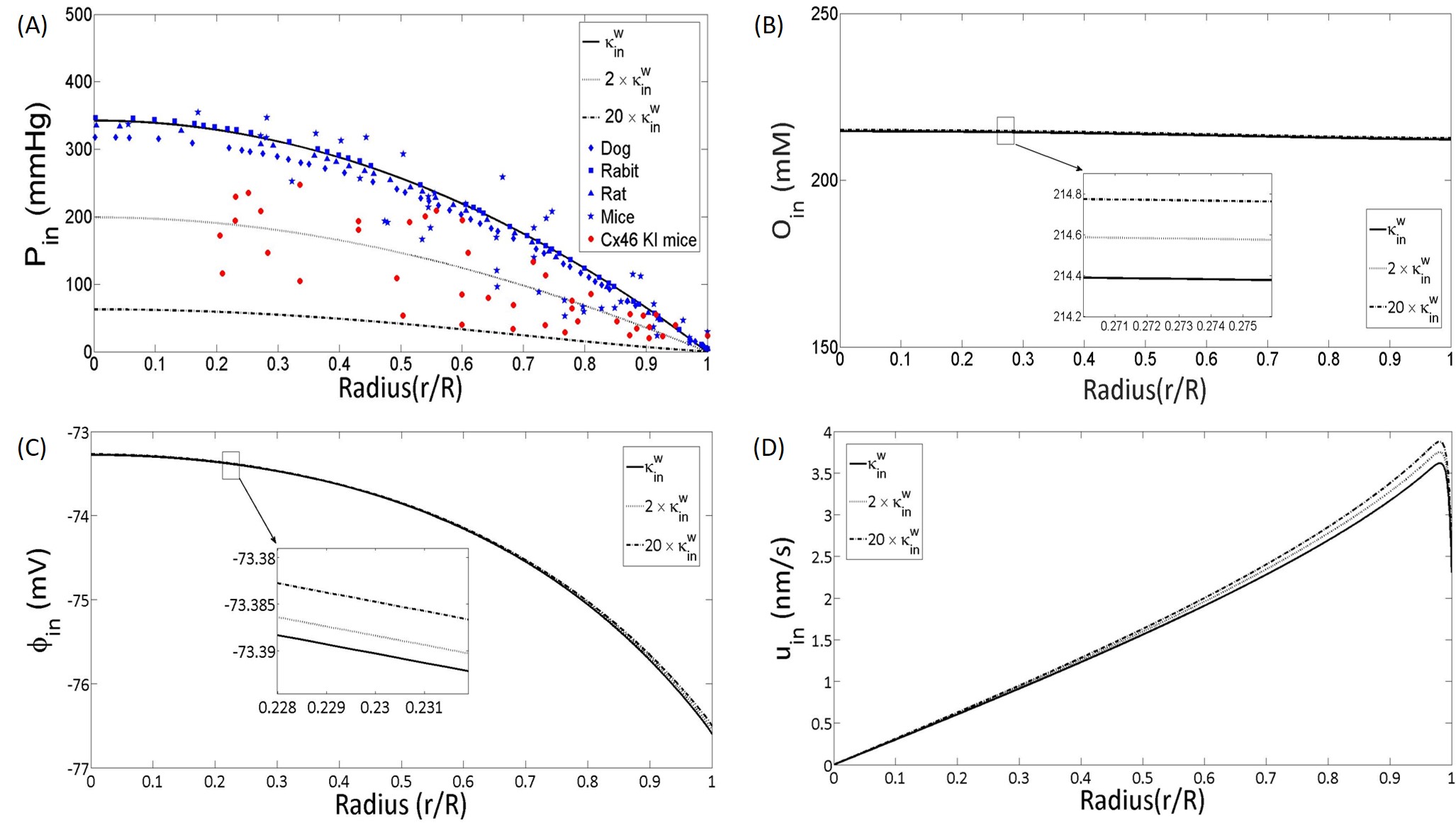}
	\caption{Comparison between different $\kappa_{in}$. The experimental data of dog, rabbit, rat come from paper \cite{RN23921}. Mice and Cx46 KI mice come from paper \cite{RN23924}. According to paper \cite{RN23924}, the Cx46 KI mice lens \textcolor{black}{has twice the number density of lens gap junction channels compared to mice}. \textcolor{black}{The parameter $\kappa_{in}^{w}=4.6830\times 10^{-20}$/ m$^{2}$} and radius \textcolor{black}{is written in dimensionless units } for different species\label{figure3}}
\end{figure*}

In Figure \ref{figure3} (A), the value  $\kappa_{in}^{w}=4.6830\times 10^{-20}/ m^{2}$ (black line) yields a good approximation to experimental data (black makers). When the conductivity of the connexins is doubled, to parameter value $\kappa_{in}$ to be $2\kappa_{in}^w$ (in the lens of mice Cx46 KI lens) as in the experiments \cite{RN23924,RN23921}, where doubled the conductivity of \textcolor{black}{the connexins} by using Cx46 KI mice lens, our model (black dot) can also match the experimental data (red markers): the intracellular hydrostatic pressure drops to half. This result shows that our full model can correctly predict the effect of permeability of membrane on hydrostatic pressure.

Interestingly, panels (B)-(D) it shows that other intracellular quantities  and extracellular ones (appendix) are insensitive to increases in the permeability by a factor of twenty, even to $20\kappa_{in}^{w}$.  The reason for this can be explained by using our simplified the \textcolor{black}{system~\ref{smodel}}. If the variation of intracellular conductance still keep the $\delta_{2}$ to be a small quantity in the dimensionless \textcolor{black}{system~\ref{fullmodelnd}}, our simplified model will be still valid. In the simplified model, All the quantities except intracellular hydrostatic pressure are related to the extracellular electric potential. However, the extracellular electric potential will not be effected by the change of the intracellular conductance, since Eq.~\ref{sphiex}  not involves intracellular conductance.  

%

\subsection{Full model vs simplified model}
\begin{figure*}[htp]
	\centering
	\includegraphics[width=6.in,height=20cm ]{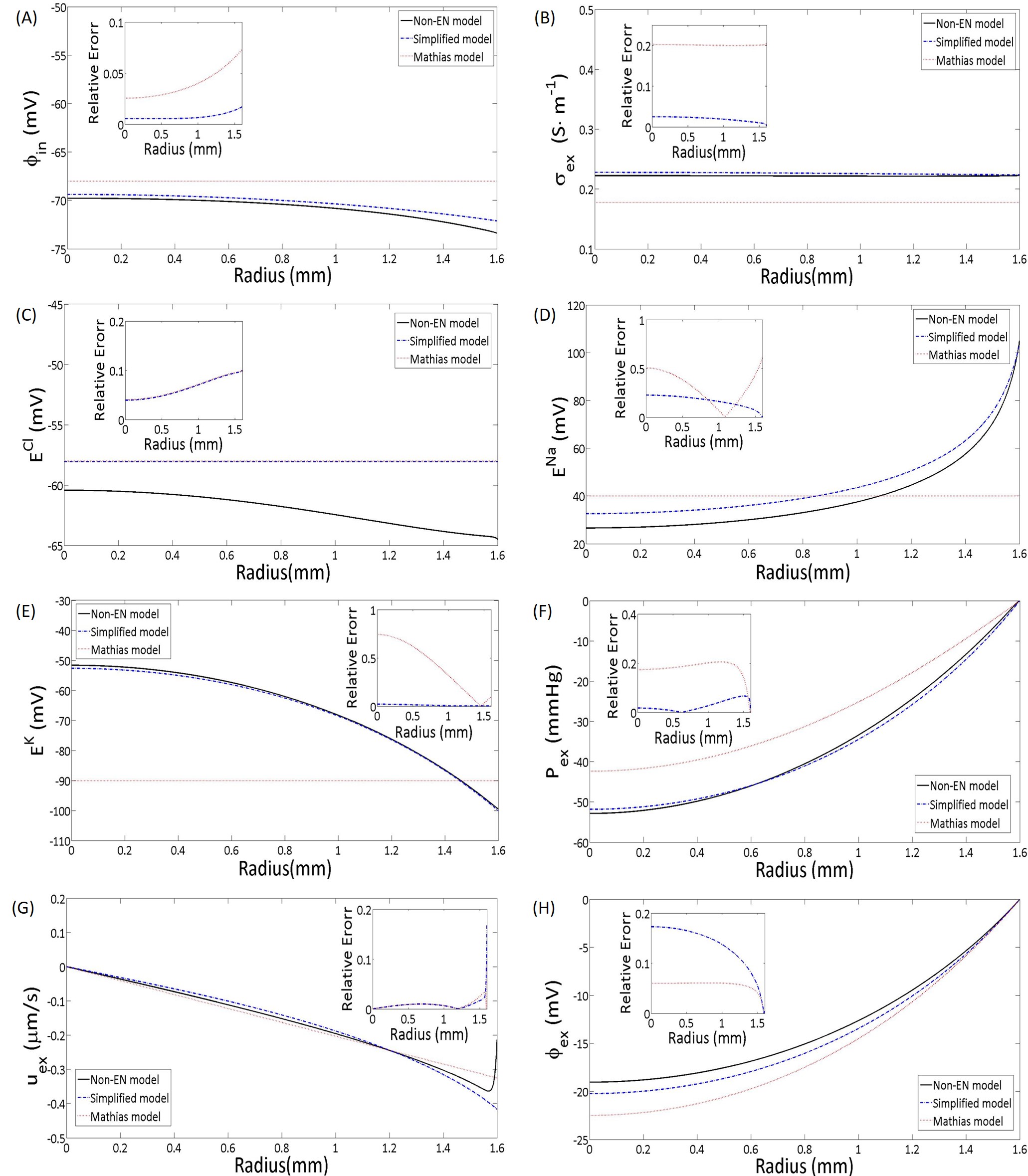}
	\caption{Comparison \textcolor{black}{of} electro-neutral and simplified and Mathias's model in \cite{RN9778}.  \label{figure4}}
\end{figure*}

In this section, we compare the full model \ref{fullmodelnd}-\ref{fullmodelndbd} with the simplified model \ref{smodel} and Mathias model in \cite{RN9778}.  The numerical results of full model (Black lines) in \textcolor{black}{Figure  \ref{figure4} (A-C)} suggest that the variations intracellular electric potential, extracellular conductance  and Nernst potential of \textcolor{black}{Cl$^-$} are  rather  small. The assumption of constant values for those variables ( potential, extracellular conductance, and Nernst, i.e. chemical potential of \textcolor{black}{Cl$^{-}$} ) in the Mathias's model (shown as red dash-dot lines) is reasonable. However, the Nernst potentials of sodium and potassium  \textcolor{black}{ (Figure  \ref{figure4} (D-E))} have large variations,  because of the effect of Sodium-potassium pump. Our simplified model (black dash lines) describes these variations with small errors. The comparisons for extracellular  pressure, velocity,  and potential  \textcolor{black}{(Figure  \ref{figure4} (F-H))} confirm that our simplified model yields good approximations to the full model.

\bigskip

\section{Conclusion}

In this paper, we propose a bidomain model to study the microcirculation  of lens. We include a capacitor in the representation of the membrane and so our model is consistent with classical electrodynamics. Consistency produces  a linear correction term in the classical charge neutrality equation.  This full model is calibrated by comparing with the experiment studying effect of connexin on hydrostatic pressure. It shows that only by changing intracellular membrane conductance (strength of connexion), our model could  match the two experimental results with different connexin very well. \textcolor{black}{Our} model is capable of making  prediction to  the circulation of lens. Furthermore, the numerical simulations show that  the  velocity, potential, osmotic pressure in the intra and extra cellular  are not sensitive to increasing conductance.    

Based on the asymptotic analysis, we proposed a simplified model, which allows us to obtain a  deep understanding  of the physical process without making unrealistic assumptions. Our results showed that the simplified model is a good  approximation of the full model where  Nernst potentials and conductivity vary significantly inside the lens. 

Our model allows calculation of variables that determine the role and life of the lens as an organ. Particularly important are the factors that determine the transparency of the lens, since that is the main function of the organ. The dependence of the size of the extracellular space, and thus the pressure in the extracellular and intracellular spaces and the difference between those two, is likely to be an important determinant of transparency. One imagines that swelling of the extracellular space will scatter light, particularly because the swelling is likely to be irregular (in a way our model does not yet capture). Changes in the Osmolarity (i.e., activity of water estimated by the total concentration of solutes) is likely to be important as well.

This hydrodynamic bidomain model can point the way to dealing with other cells, tissues, and organs in which current flow, water flow, and cell volume changes are important. These include the kidney, the central nervous system (where the narrow extracellular space poses many of the biological problems facing the lens), the t-tubular system of skeletal and much cardiac muscle and so on. We show that a mathematically well defined model can deal with the reality of biological structure and its complex distribution of channels, etc. 

\textcolor{black}{Conservation laws applied to simplified structures are enough to provide quite useful results, as they were in three dimensional electrical problems of cells of various geometries \cite{RN575}
	and syncytia \cite{RN25704}-\cite{RN918}. The exact results are analyzed with perturbation methods, described in general in \cite{peskoff1973interpretation} 
	and these methods allow dramatic simplifications without introducing large or even significant errors. It is as if evolution chose systems in which parameters and structures allow simple results, in which parameters can control biological function robustly.}

Of course, we only point the way. Additional compartments and additional structural complexity will surely be needed to deal with the workings of evolution. But these can be handled in a mathematically defined way, yielding approximate results with clear physical and biological interpretation. Combining the multi-domain model and membrane potential dependent conductance, \textcolor{black}{one can model depolarization induced by extra potassium in lens  \cite{delamere1977comparison,malcolm2006computational}
	and cortical spreading depression (CSD) problem \cite{Joshua2013,Yao2011,MORI201594}. 
	The ultimate goals will be (i) to provide as much precision in the mathematics and physics as we can, starting from first principles \cite{RN28366}; (ii) to provide a general basis for treatments of convection in other tissues that involve microcirculation. Computational models of these are not in hand, and may be hard to construct, since so little is know of those systems compared to the lens. With what we have learned here, we hope a general mathematical approach and model of the type we present here may be constructed and helpful in other systems with narrow extracellular spaces that are likely to need microcirculation to augment diffusion, like cardiac and skeletal muscle, kidney, liver, epithelia, and the extracellular space of the brain.} \\ 

\noindent\textbf{Author Contributions}

\textcolor{black}{Y.Z, S.X, and H.H did the model derivations and carried out the numerical simulations. R.S.E and H.H designed the study, coordinated the study, and commented on the manuscript. All authors gave final approval for publication.
}

\noindent\textbf{Acknowledgments}

This research is supported in part by the Fields Institute for Research in Mathematical Science (S.X., \textcolor{black}{R.S.E.}, H.H.), the Natural Sciences and Engineering Research Council of Canada (H.H.).

%
  
\section*{References}
\normalem
\bibliography{axivversion}

\appendix
\section{Model Parameters}\label{AppendixA}
\begin{center}
	\begin{scriptsize}
		\begin{tabular}{ |p{1.5cm}|p{2cm}|p{2.5cm}|p{1.5cm}|p{2.5cm}|p{2.5cm}|  }
			\hline
			Parameters & Mathias \cite{RN9778} & Malcolm \cite{malcolm2006computational}&Parameters & Mathias \cite{RN9778} & Malcolm \cite{malcolm2006computational}\\
			\hline
			\hline
			$R$       &  $1.6\times 10^{-3}$ m &$1.6\times 10^{-3}$ m & $L_{m}$  & $3.75\times10^{-13}m /(Pa\cdot s)$ & $1.34\times10^{-13}m /(Pa\cdot s)$\\
			$A_{in}/V_{in}$   & $78$ mM &$78$ mM& $L_{s}$ &$3.75\times10^{-13}m /(Pa\cdot s)$& $8.89\times10^{-13}m /(Pa\cdot s)$ \\
			$C^{Na}_{o}$  &  $107$ mM &$107$ mM &  $\mathcal{M}_{in}$& $ 0.988 $& $0.99$\\
			$C^{K}_{o}$   &  $3$ mM   & $3$ mM  &   $\mathcal{M}_{ex}$& $0.012$& $0.01$\\
			$C_{m}$   &  - & $1\times 10^{-2}\ F/m^{2}$ & $\mathcal{M}_{v}$&$6\times 10^{5}/m$&$5\times 10^{5}/m$\\
			$D^{Na}_{ex}$ & -  & $1.39 \times 10^{-9}\ m^{2}/s$ &$T$& -&$310\ K$\\
			$D^{K}_{ex}$  & -  & $2.04\times 10^{-9} \ m^{2}/s$& $k_{e}$&$1.72\times 10^{-8}\ m^{2}/(V\cdot s)$& $1.45\times 10^{-8}\ m^{2}/(V\cdot s)$\\
			$D^{Cl}_{ex}$ & -  &  $2.12\times 10^{-9} \ m^{2}/s$& $k_{B}$&$1.38\times 10^{-23}\ J/K$&$1.38\times 10^{-23}\ J/K$\\
			$D^{Na}_{in}$ & -  & $1.39\times 10^{-11} \ m^{2}/s$&$K_{K1}$&-&$1.6154\ mM$\\
			$D^{K}_{in}$  & -  & $2.04\times 10^{-11}\ m^{2}/s$ &$K_{K2}$&-&$ 0.1657\ mM$\\
			$D^{Cl}_{in}$ & -  &$2.12\times 10^{-11} \ m^{2}/s$&$K_{Na1,Na2}$&-&$ 2.3393\ mM$\\
			$e$ & $1.6\times 10^{-19}\ A\cdot s$  & $1.6\times 10^{-19} A\cdot s$&$\eta$&$0.988$&$0.99$\\
			$g^{Na}$      & $2.2\times10^{-3} \ S/m^{2}$  & $2.2 \times 10^{-3} S/m^{2}$&$\kappa_{ex}$&$1.141\times 10^{-16}\ m^{2}$&$1.33\times 10^{-16}\ m^{2}$\\
			$g^{Cl}$      & $2.2 \times 10^{-3} \ S/m^{2}$ & $2.2 \times 10^{-3}\ S/m^{2}$ & $\kappa_{in}$&- &$9.366\times10^{-19}\ m^{2}$ \\
			$G^{K}$       &  $2.1 \ S/m^{2}$ &$2.1\ S/m^{2}$& $\gamma_{m,s}$&1&1\\
			$I_{p}$       &  $2.3 \times 10^{-2} A/m^{2}$ &- &$\tau_{c}$&0.16&0.16\\
			$I_{max1}$    & -  &$0.478 \  A/m^{2} $ &$\mu$&$7\times 10^{-4}\ Pa\cdot s$&$7\times 10^{-4}\ Pa\cdot s$\\
			$I_{max2}$    & -  &$0.065 \ A/m^{2}  $ & $\bar{z}$&-1.5&-1.5\\
			\hline
		\end{tabular}
	\end{scriptsize}
\end{center}	

\newpage
\section{Dimensionless Parameters and Scales}\label{AppendixB}	
The following dimensionless parameters' value and scales calculation based on  values in \cite{malcolm2006computational}
\begin{center}
	\begin{scriptsize}
		\begin{tabular}{ |p{2cm}|p{2.5cm}|p{2cm}|p{2.5cm}|  }
			\hline
			Scales/Parameters & Value & Parameters & Value\\
			\hline
			\hline
			$a^{Na*}$& $6.9 \times 10^{-2} A/m^{2}$&$\delta_{0}=\frac{\mathcal{M}_{ex}}{\mathcal{M}_{in}}$& $\frac{1}{99}$\\	
			$C^{*}$  & $110\ mM$&	 $\delta_{1}=\frac{k_{e}\tau_{c} k_{B}T}{e R u^{*}_{ex}}$& $1.2031\times 10^{-1}$\\		
			$O^{*}$  & $220\ mM$ &$\delta_{2}=\frac{\mu R u^{*}_{in}}{\kappa_{in}\gamma_{m} k_{B}T O^{*}}$& $6.861\times 10^{-3} $\\	
			$P^{*}$  & $16.937 \ KPa$&$\delta_{3}=\frac{P^{*}}{\gamma_{m} k_{B}T O^{*}} $& $2.9894\times 10^{-2}$\\	
			$u_{in}^{*}$  & $3.2506 \ nm/s$ &$\delta_{4}=\frac{\mathcal{M}_{in}u^{*}_{in}}{R\mathcal{M}_{v}L_{m}\gamma_{m}k_{B}TO^{*}}$& $3.5323\times 10^{-5}$\\		 
			$u_{ex}^{*}$  & $3.2181\ \mu m/s$&$\delta_{5}=\frac{u^{*}_{in}}{L_{s}\gamma_{s}k_{B}TO^{*}}$&$4.3022\times 10^{-3}$\\
			$\phi^{*}$  & $26.7\ mV$&$\delta_{6}=\frac{\mathcal{M}_{v}C_{m}k_{B}T}{e^{2}C^{*}\eta}$& $1.2745\times 10^{-5}$\\
			$D^{*}_{ex}$  & $3.392 \times 10^{10}\ m^{2}/s$& $\delta_{7}=\frac{\mathcal{M}_{v}C_{m}k_{B}T}{e^{2}C^{*}(1-\eta)}$& $1.2617\times 10^{-3}$\\
			$D^{*}_{in}$  & $2.12\times 10^{-11}\ m^{2}/s$&$\delta_{8}=\frac{\mathcal{M}_{ex}D^{*}_{ex}}{\mathcal{M}_{in}D^{*}_{in}}$& $1.6162\times 10^{-1}$\\
			$Pe_{ex}$  & $1.5180$&$\delta_{9}=\frac{D^{Cl}_{l}-D^{K}_{l}}{D^{*}_{l}}$&$3.77\times 10^{-2}$ \\
			$Pe_{in}$  & $2.4533\times 10^{-1}$&$\delta_{10}=\frac{D^{Cl}_{l}-D^{Na}_{l}}{D^{*}_{l}}$& $3.443\times 10^{-1}$\\
			$\widetilde{D}^{Na}_{in,ex}$  & $0.6557$& $\delta_{11}=\frac{C_{in}^{Cl,0}}{C_{ex}^{Cl,0}}$ & $\frac{12.5}{110}$\\
			$\widetilde{D}^{K}_{in,ex}$  & $0.9623$& $\rho_{0}$& $\frac{117}{110}$\\
			$\widetilde{D}^{Cl}_{in,ex}$  & $1$& $\widetilde{\mathcal{M}}^{in}_{v}$ & $3.3859\times 10^{-1}$\\
			$R_{s}$  & $4.00\times 10^{-1}$&$\widetilde{\mathcal{M}}^{ex}_{v}$ & $2.095$\\
			\hline
		\end{tabular}
	\end{scriptsize}
\end{center}	
The $\delta$ can be find in the following equations. 
\begin{multicols}{2}
	\begin{equation*}
	\begin{aligned}
	&\delta_{0}:\ in\ eq.[13],\\
	&\delta_{1}:\ in\ eq.[14a],\\
	&\delta_{2}:\ in\ eq.[14b],\\
	&\delta_{3}:\ in\ eq.[14b],\\
	&\delta_{4}:\ in\ eq.[14c],\\
	&\delta_{5}:\ in\ B.C.\ below\ eq.[14],\\
	& \delta_{6}:\ in\ eq.[14e],\\
	\end{aligned}
	\end{equation*}
	\columnbreak
	
	\begin{equation*}
	\begin{aligned}
	&\delta_{7}:\ in\ eq.[14f],\\
	&\delta_{8}:\ in\ eq.[14h],\\
	&\delta_{9}:\ in\ eq.[18],\\
	&\delta_{10}:\ in\ eq.[18],\\
	&\delta_{11}:\ in\ eq.[43],\\
	\end{aligned}
	\end{equation*}
\end{multicols}
\newpage	
\section{Non-dimensionalization}\label{Dimensionless systems}
In this section, we derive the dimensionless model based on the lens, which has been widely studied. The major ions we considering here are sodium ($Na^{+}$), potassium ($K^{+}$) and chloride $(Cl^{-})$ and the sodium-potassium pump which distributed on the surface of the lens. Although we restrict ourselves in this particular problem, the following procedure can be applied in a wide range of practical problems in biological syncytia. 
\subsection{Water circulation}
In the following, we assume the typical length of lens is $R$. The fluid system is driven by the osmotic gradient, which is generated by the sodium-potassium pump on the surface.
In Eq. 7, the strength of sodium-potassium pump at surface depends on the ion's concentration , which leads
\begin{equation}
\label{PP3}
a^{Na}=3\frac{I_{p}}{e}, \ \ \  a^{K}=-2\frac{I_{p}}{e}, \ \ \
a^{Cl}=0,
\end{equation}
where 
\begin{equation}
\label{PP4}
\begin{aligned}
I_{p}=I_{max1}\left(\frac{C^{Na}_{in}}{C^{Na}_{in}+K_{Na1}}\right)^{3}\left(\frac{C^{K}_{o}}{C^{K}_{o}+K_{K1}}\right)^{2}+I_{max2}\left(\frac{C^{Na}_{in}}{C^{Na}_{in}+K_{Na2}}\right)^{3}\left(\frac{C^{K}_{o}}{C^{K}_{o}+K_{K2}}\right)^{2}.
\end{aligned}
\end{equation}
We assume that the velocity at surface determines the characteristic velocity scale for the problem. We have ion fluxes in the intracellular, extracellular region in Eq. 5  and trans-membrane source of ion in  Eq. 6 for ion $Na^{+},K^{+},Cl^{-}$.\\
At boundary of the intracellular space, due to the ion pump in Eq. \ref{PP3}  and assumption of conductance at surface that $G^{Na}=G^{Cl}=0$ \cite{RN9778,RN23921}, we have 
\begin{equation}
\label{BFL1}
\begin{aligned}
&J^{Na}_{in}=a^{Na},\ \ \ J^{K}_{in}=j^{K}_{s}+a^{K},\ \ \ J^{Cl}_{in}=0.
\end{aligned}
\end{equation}
Since $g^{K}=0$ inside of the lens, we obtain 
\begin{equation}
\label{KB}
j^{K}_{s}+a^{K}=0.
\end{equation}
This assumption obviously will have to be replaced in applications to other tissues, with a less particular distribution of channel proteins.\\
By the conservation of fluxes for each ion in Eq. 4, we get
\begin{equation}
J^{i}_{in}=-\delta_{0}J^{i}_{ex}, \ \ \ \ i=Na,K,Cl,
\end{equation}
where $\delta_{0}=\frac{\mathcal{M}_{ex}}{\mathcal{M}_{in}}$. Therefore, Eq. \ref{BFL1}  becomes
\begin{equation}
\label{bdf}
\begin{aligned}
-\delta_{0}J^{Na}_{ex}=a^{Na},\ \ \ -\delta_{0}J^{K}_{ex}=0,\ \ \ -\delta_{0}J^{Cl}_{ex}=0.
\end{aligned}
\end{equation}
Adding up all three fluxes in Eq. \ref{bdf}  and since in the extracellular region each ion diffusion coefficient are at the same level of approximation, i.e.
\begin{equation}
D^{i}_{ex}=O \left(D_{ex} \right), \ \ \ i=Na,K,Cl,
\end{equation}
and based on  Eq. 10 , we get
\begin{equation}
\label{SC2}
O_{ex}u_{in}+\delta_{0}D_{ex}\tau_{c}\frac{d}{dr} O_{ex}+\frac{D_{ex}\tau_{c}\delta_{0}}{k_{B}T}\rho_{ex}\frac{d}{dr} \phi_{ex}=a^{Na}.
\end{equation}
The strength of the ion pump $a^{Na}$ depends on the ion concentration in Eq. \ref{PP4} . We choose the scale of $a^{Na}$ is $a^{Na*}$ based on an experimental estimation \cite{RN9778}. Using Eq. \ref{SC2}, we take the scale for $O_{in,ex}$ and $u_{in}$ to be $O^{*}$ and $u^{*}_{in}$ as 
\begin{equation}
\label{SC3}
\begin{aligned}
O^{*}=2\left(C^{Na}_{o}+C^{K}_{o}\right),\ \ \ u^{*}_{in}=\frac{a^{Na*}}{O^{*}}.
\end{aligned}
\end{equation}
By mass conservation expressed in  Eq. 1, we naturally get the scale of $u_{ex}$ as
\begin{equation}
\label{UeS}
u^{*}_{ex}=\delta_{0}^{-1} u^{*}_{in}.
\end{equation}
Furthermore, $\phi^{*}=\frac{k_{B}T}{e}$ is used for the scale of electric potential $\phi_{in}$ and $\phi_{ex}$. For the extracellular velocity in Eq. 2, we have
\begin{equation}
u^{*}_{ex}\widetilde{u}_{ex}=-\frac{\kappa_{ex}}{\mu R}\tau_{c}P^{*}_{ex}\frac{d}{d \widetilde{r} } \widetilde{P}_{ex}-k_{e}\tau_{c}\frac{k_{B}T}{e R}\frac{d}{d \widetilde{r} } \widetilde{\phi}_{ex},
\end{equation}
We think the $\frac{d}{d r } P_{ex}$ term balance the velocity $u_{ex}$. The scale for extracellular pressure $P^{*}_{ex}$ is then choose
\begin{equation*}
P^{*}_{ex}=\frac{\mu R u^{*}_{ex}}{\kappa_{ex}\tau_{c}}.
\end{equation*}
Therefore, we get
\begin{equation}
\label{u_ex}
\widetilde{u}_{ex}=-\frac{d}{d \widetilde{r} } \widetilde{P}_{ex}-\delta_{1}\frac{d}{d \widetilde{r} }\widetilde{\phi}_{ex},
\end{equation}
where $\delta_{1}=\frac{k_{e}\tau_{c} k_{B}T}{e R u^{*}_{ex}}$.
For the intracellular velocity, we have
\begin{equation}
\label{Uin1}
u^{*}_{in}\widetilde{u}_{in}=-\frac{\kappa_{in} P^{*}_{in}}{\mu R} \frac{d}{d \widetilde{r} } \widetilde{P}_{in}+\frac{\kappa_{in}\gamma_{m} k_{B}T O^{*}}{\mu R}\frac{d}{d \widetilde{r} } \widetilde{O}_{in}.
\end{equation}
We claim term $\frac{d}{d r } P_{in}$ and $\frac{d}{d r } O_{in}$ balance at the same level. Therefore, we choose the same scale for the intracellular and extracellular pressure, namely,
\begin{equation*}
P^{*}=P^{*}_{in}=P_{ex}^{*}.
\end{equation*}
Then Eq. \ref{Uin1}  becomes
\begin{equation}
\label{u_in}
\delta_{2}\widetilde{u}_{in}=-\delta_{3}\frac{d}{d \widetilde{r} } \widetilde{P}_{in}+\frac{d}{d \widetilde{r} } \widetilde{O}_{in},
\end{equation}
where 
\begin{equation*}
\begin{aligned}
\delta_{2}=\frac{\mu R u^{*}_{in}}{\kappa_{in}\gamma_{m} k_{B}T O^{*}},
\ \ \ \ \delta_{3}=\frac{P^{*}}{\gamma_{m} k_{B}T O^{*}}.
\end{aligned}
\end{equation*}
In all, the fluid system Eq. 1 becomes
\begin{equation}
\label{FS}
\left\{
\begin{aligned}
&\widetilde{u}_{ex}=-\widetilde{u}_{in},\\
&\delta_{4}\frac{1}{\widetilde{r}^2}\frac{d}{d \widetilde{r}} \left(\widetilde{r}^2  \widetilde{u}_{in}\right)=\delta_{3} \left(\widetilde{P}_{ex}-\widetilde{P}_{in}\right)+\left(\widetilde{O}_{in}-\widetilde{O}_{ex}\right),\\
\end{aligned}
\right.
\end{equation}
with boundary condition
\begin{equation*}
\left\{
\begin{aligned}
&\widetilde{P}_{ex}=0, \\
&\delta_{5}\widetilde{u}_{in}=\delta_{3} \widetilde{P}_{in}-\left(\widetilde{O}_{in}-\widetilde{O}_{ex}\right),
\end{aligned}
\right.
\end{equation*}
where
\begin{equation*}
\begin{aligned}
\delta_{4}=\frac{\mathcal{M}_{in}u^{*}_{in}}{R\mathcal{M}_{v}L_{m}\gamma_{m}k_{B}TO^{*}},\ \ \ 
\delta_{5}=\frac{u^{*}_{in}}{L_{s}\gamma_{s}k_{B}TO^{*}}.
\end{aligned}
\end{equation*}
\subsection{Ions circulation}
The velocity scales and diffusion coefficients in the extracellular and intracellular space are at different levels of approximation in our approach. In the following, we put the characteristic diffusion coefficients at intracellular and extracellular region  and  scale of concentration as 
\begin{equation*}
D^{*}_{ex}=D^{Cl}_{ex}\tau_{c},\ \ \ 
D^{*}_{in}=D^{Cl}_{in} , \ \ \  C^{*}=C^{Na}_{o}+C^{K}_{o}.
\end{equation*}
In this way, we get Peclet number in the extracellular and intracellular  and dimensionless Nernst potential as  
\begin{equation*}
Pe_{in}=\frac{u^{*}_{in} R}{D^{*}_{in}},\ \ \
Pe_{ex}=\frac{u^{*}_{ex} R}{D^{*}_{ex}}, \ \ \
\widetilde{E}^{i}=\frac{1}{z^{i}}\log\left(\frac{\widetilde{C}^{i}_{ex}}{\widetilde{C}^{i}_{in}}\right).
\end{equation*}
Because  $g^{Na} = 0$ inside of lens, we have $K^{+}$ system as in Mathias's model \cite{RN9778},
\begin{equation}
\label{Ks}
\left\{
\begin{aligned}
& \frac{1}{\widetilde{r}^2}\frac{d}{d \widetilde{r}} \left(\widetilde{r}^2 \left( Pe_{ex} \widetilde{C}^{K}_{ex}\widetilde{u}_{ex}-\widetilde{D}^{K}_{ex} \left(\frac{d}{d \widetilde{r} } \widetilde{C}^{K}_{ex}+z^{K}\widetilde{C}^{K}_{ex}\frac{d}{d \widetilde{r} }\widetilde{\phi}_{ex} \right) \right) \right)=0,\\
&\frac{1}{\widetilde{r}^2}\frac{d}{d \widetilde{r}} \left(\widetilde{r}^2  \left(Pe_{in}  \widetilde{C}^{K}_{in}\widetilde{u}_{in}-\widetilde{D}^{K}_{in}\left(\frac{d}{d \widetilde{r} } \widetilde{C}^{K}_{in}+z^{K}\widetilde{C}^{K}_{in} \frac{d}{d \widetilde{r} }\widetilde{\phi}_{in} \right) \right)\right)=0,\\
\end{aligned}
\right.
\end{equation}
with boundary condition
\begin{equation*}
\left\{
\begin{aligned}
&\widetilde{C}^{K}_{ex}=\widetilde{C}^{K}_{o}, \\
& Pe_{in}  \widetilde{C}^{K}_{in}\widetilde{u}_{in}-\widetilde{D}^{K}_{in}\left(\frac{d}{d \widetilde{r} } \widetilde{C}^{K}_{in}+z^{K}\widetilde{C}^{K}_{in} \frac{d}{d \widetilde{r} } \widetilde{\phi}_{in} \right) =\frac{R_{s}}{z^{K}}\left(\widetilde{\phi}_{in}-\widetilde{E}^{K}\right)+\widetilde{a}^{K},
\end{aligned}
\right.
\end{equation*}
and $Cl^{-}$ system as
\begin{equation}
\label{Cls}
\left\{
\begin{aligned}
&\frac{1}{\widetilde{r}^2}\frac{d}{d \widetilde{r}} \left(\widetilde{r}^2\left(Pe_{ex}  \widetilde{C}^{Cl}_{ex}\widetilde{u}_{ex}-\widetilde{D}^{Cl}_{ex}\left(\frac{d}{d \widetilde{r} } \widetilde{C}^{Cl}_{ex}+z^{Cl}\widetilde{C}^{Cl}_{ex} \frac{d}{d \widetilde{r} }\widetilde{\phi}_{ex} \right) \right)\right)=\frac{ \widetilde{\mathcal{M}}_{v}^{ex}}{z^{Cl}}\left( \widetilde{\phi}_{in}-\widetilde{\phi}_{ex}-\widetilde{E}^{Cl}\right) ,\\
&\frac{1}{\widetilde{r}^2}\frac{d}{d \widetilde{r}} \left(\widetilde{r}^2\left(Pe_{in}  \widetilde{C}^{Cl}_{in}\widetilde{u}_{in}-\widetilde{D}^{Cl}_{in}\left(\frac{d}{d \widetilde{r} } \widetilde{C}^{Cl}_{in}+z^{Cl}\widetilde{C}^{Cl}_{in} \frac{d}{d \widetilde{r} } \widetilde{\phi}_{in} \right) \right)\right)\\
&~~= -\delta_{8} \frac{1}{\widetilde{r}^2}\frac{d}{d \widetilde{r}} \left(\widetilde{r}^2 \left(Pe_{ex} \widetilde{C}^{Cl}_{ex}\widetilde{u}_{ex}-\widetilde{D}^{Cl}_{ex}\left(\frac{d}{d \widetilde{r} } \widetilde{C}^{Cl}_{ex}+z^{Cl}\widetilde{C}^{Cl}_{ex}\frac{d}{d \widetilde{r} } \widetilde{\phi}_{ex} \right) \right)\right),\\
\end{aligned}
\right.
\end{equation}
with boundary condition
\begin{equation*}
\left\{
\begin{aligned}
& \widetilde{C}^{Cl}_{ex}=\widetilde{C}^{Na}_{o}+\widetilde{C}^{K}_{o}+\delta_{7}\left(\widetilde{\phi}_{in}-\widetilde{\phi}_{ex}\right),\\
&Pe_{in} \widetilde{C}^{Cl}_{in}\widetilde{u}_{in}-\widetilde{D}^{Cl}_{in}\left(\frac{d}{d \widetilde{r} } \widetilde{C}^{Cl}_{in}+z^{Cl}\widetilde{C}^{Cl}_{in} \frac{d}{d \widetilde{r} }\widetilde{\phi}_{in} \right) =0.
\end{aligned}
\right.
\end{equation*}
where 
\begin{equation*}
R_{s}=\frac{G^{K}k_{B}TR}{e^{2}D^{*}_{in}C^{*}},\ \ \ 
\widetilde{a}^{K}=\frac{a^{K}R}{D^{*}_{in}C^{*}},\ \ \
\widetilde{\mathcal{M}}_{v}^{ex}=\frac{\mathcal{M}_{v}g^{Cl}k_{B}TR^{2}}{\mathcal{M}_{ex}e^{2}D^{*}_{ex}C^{*}}, \ \ \ \delta_{8}=\frac{\mathcal{M}_{ex}D^{*}_{ex}}{\mathcal{M}_{in}D^{*}_{in}}.
\end{equation*}	
The concentration of $Na^{+}$ can be solved from the following equations 
\begin{equation}
\label{nonEN2}
\left\{
\begin{aligned}
&\sum_{i}z^{i}\widetilde{C}^{i}_{in}+\bar{z}\widetilde{\frac{A_{in}}{V_{in}}}=\delta_{6}\left(\widetilde{\phi}_{in}-\widetilde{\phi}_{ex}\right),\\
&\sum_{i}z^{i}\widetilde{C}^{i}_{ex}=-\delta_{7}\left(\widetilde{\phi}_{in}-\widetilde{\phi}_{ex}\right),\\
\end{aligned}
\right.
\end{equation}
where
\begin{equation}
\delta_{6}=\frac{\mathcal{M}_{v}C_{m}k_{B}T}{e^{2}C^{*}\eta},\ \ \
\delta_{7}=\frac{\mathcal{M}_{v}C_{m}k_{B}T}{e^{2}C^{*}(1-\eta)}.
\end{equation}

From Eq. 11 and use the fact $z^{Na}=z^{K}=1$ and assumption that  $g^{Na}=g^{Cl}$ and $G^{Na}=G^{Cl}=0$, we have
\begin{equation}
\label{potens}
\left\{
\begin{aligned}
&\frac{1}{\widetilde{r}^2}\frac{d}{d \widetilde{r}} \left(\widetilde{r}^2 \left( Pe_{ex}\widetilde{\rho}_{ex}\widetilde{u}_{ex}-\sum_{i} \widetilde{D}^{i}_{ex}z^{i}\frac{d}{d \widetilde{r} }\widetilde{C}^{i}_{ex}-\widetilde{\sigma}_{ex}\frac{d}{d \widetilde{r} } \widetilde{\phi}_{ex}\right)\right)=\widetilde{\mathcal{M}}_{v}^{ex}\left(2\left(\widetilde{\phi}_{in}-\widetilde{\phi}_{ex}\right)-\widetilde{E}^{Na}-\widetilde{E}^{Cl}\right),\\
&\frac{1}{\widetilde{r}^2}\frac{d}{d \widetilde{r}} \left(\widetilde{r}^2  \left( Pe_{in} \widetilde{\rho}_{in} \widetilde{u}_{in}-\sum_{i}\widetilde{D}^{i}_{in}z^{i}\frac{d}{d \widetilde{r} }\widetilde{C}^{i}_{in}-\widetilde{\sigma}_{in}\frac{d}{d \widetilde{r} }\widetilde{\phi}_{in}\right)\right) \\
&~~=-\delta_{8}\frac{1}{\widetilde{r}^2}\frac{d}{d \widetilde{r}} \left(\widetilde{r}^2  \left( Pe_{ex}\widetilde{\rho}_{ex}\widetilde{u}_{ex}-\sum_{i} \widetilde{D}^{i}_{ex}z^{i}\frac{d}{d \widetilde{r} }\widetilde{C}^{i}_{ex}-\widetilde{\sigma}_{ex}\frac{d}{d \widetilde{r} } \widetilde{\phi}_{ex}\right)\right),\\
\end{aligned}
\right.
\end{equation}

with boundary condition
\begin{equation}
\left\{
\begin{aligned}
&\widetilde{\phi}_{ex}=0,\\
&Pe_{in} \widetilde{\rho}_{in} \widetilde{u}_{in}-\sum_{i}\widetilde{D}^{i}_{in}z^{i}\frac{d}{d \widetilde{r} }\widetilde{C}^{i}_{in}-\widetilde{\sigma}_{in}\frac{d}{d \widetilde{r} }\widetilde{\phi}_{in} =\frac{R_{s}}{z^{K}}\left(\widetilde{\phi}_{in}-\widetilde{E}^{K}\right)+\widetilde{I}^{\phi}_{p},
\end{aligned}
\right.
\end{equation}
where 
\begin{equation*}
\widetilde{\rho}_{in}=|\bar{z}|\widetilde{\frac{A_{in}}{V_{in}}}+\delta_{6} \left( \widetilde{\phi}_{in}-\widetilde{\phi}_{ex}\right),\ \
\widetilde{\rho}_{ex}=\delta_{7} \left(\widetilde{\phi}_{ex}-\widetilde{\phi}_{in} \right),\ \ \ 
\widetilde{I}^{\phi}_{p}=\frac{I_{p}R}{eD^{*}_{in}C^{*}},
\end{equation*}
and 
\begin{equation*}
\label{conduc}
\widetilde{\sigma}_{in}=\sum_{i} \widetilde{D}^{i}_{in}(z^{i})^{2}\widetilde{C}^{i}_{in},  \ \ \ \  \ \ \widetilde{\sigma}_{ex}=\sum_{i} \widetilde{D}^{i}_{ex}(z^{i})^{2}\widetilde{C}^{i}_{ex}.
\end{equation*}

%

\section{Effect of permeability}\label{Permeability changed}
\begin{figure}[H]
	\centering
	\begin{minipage}[c]{0.45\textwidth}
		\centering
		\includegraphics[width=3.5in,height=2.16in]{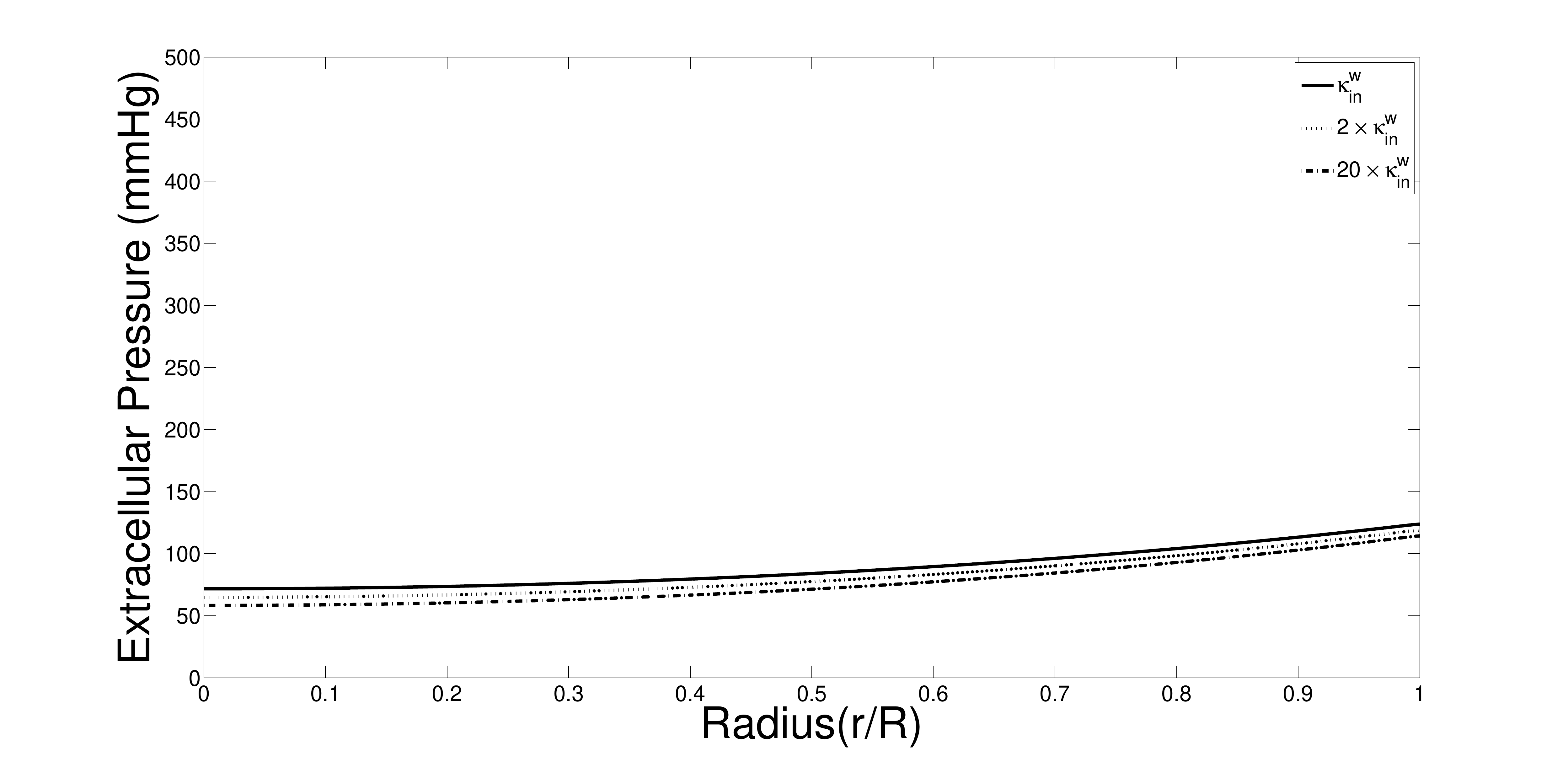}
	\end{minipage}
	\hfill\begin{minipage}[c]{0.45\textwidth}
		\centering
		\includegraphics[width=3.5in,height=2.16in]{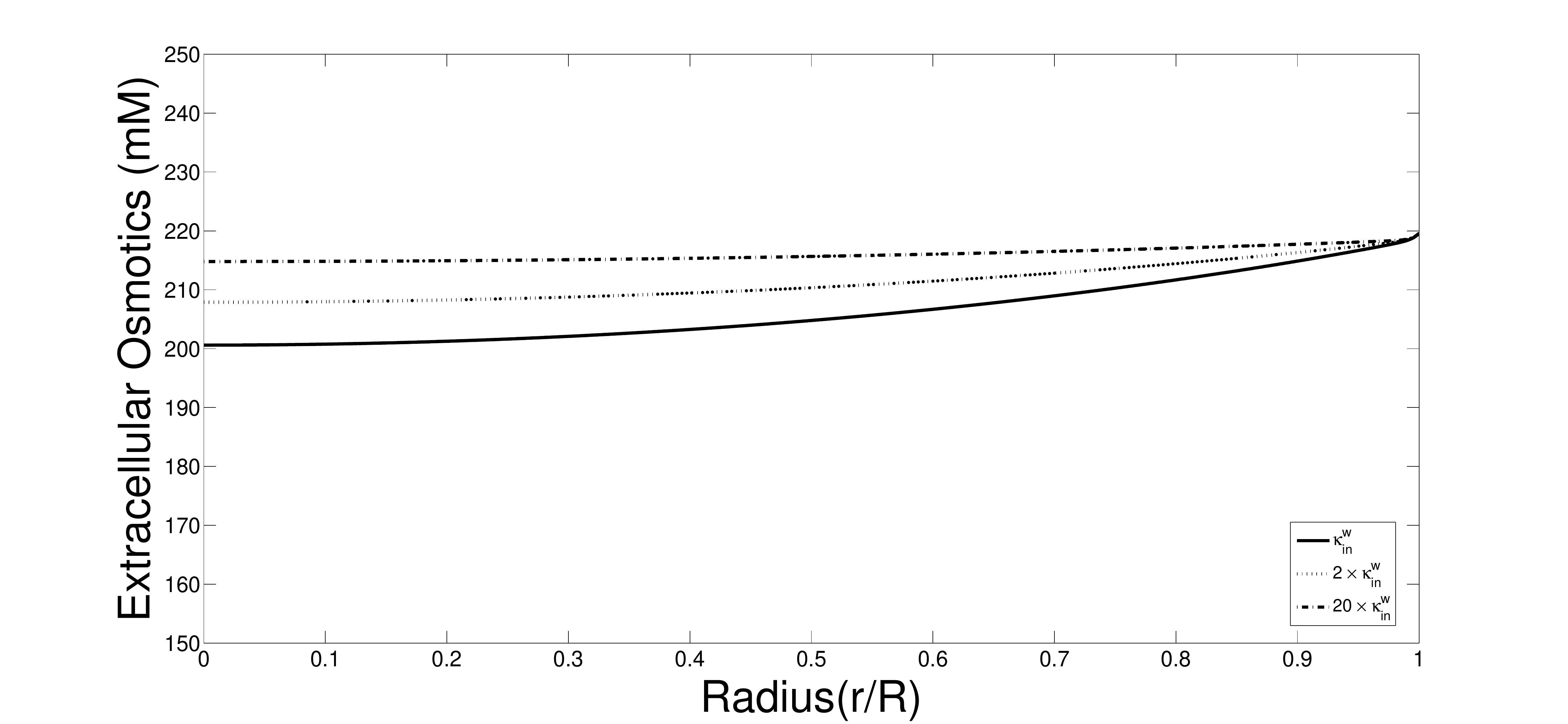}
	\end{minipage}
	\begin{minipage}[c]{0.45\textwidth}
		\centering
		\includegraphics[width=3.5in,height=2.16in]{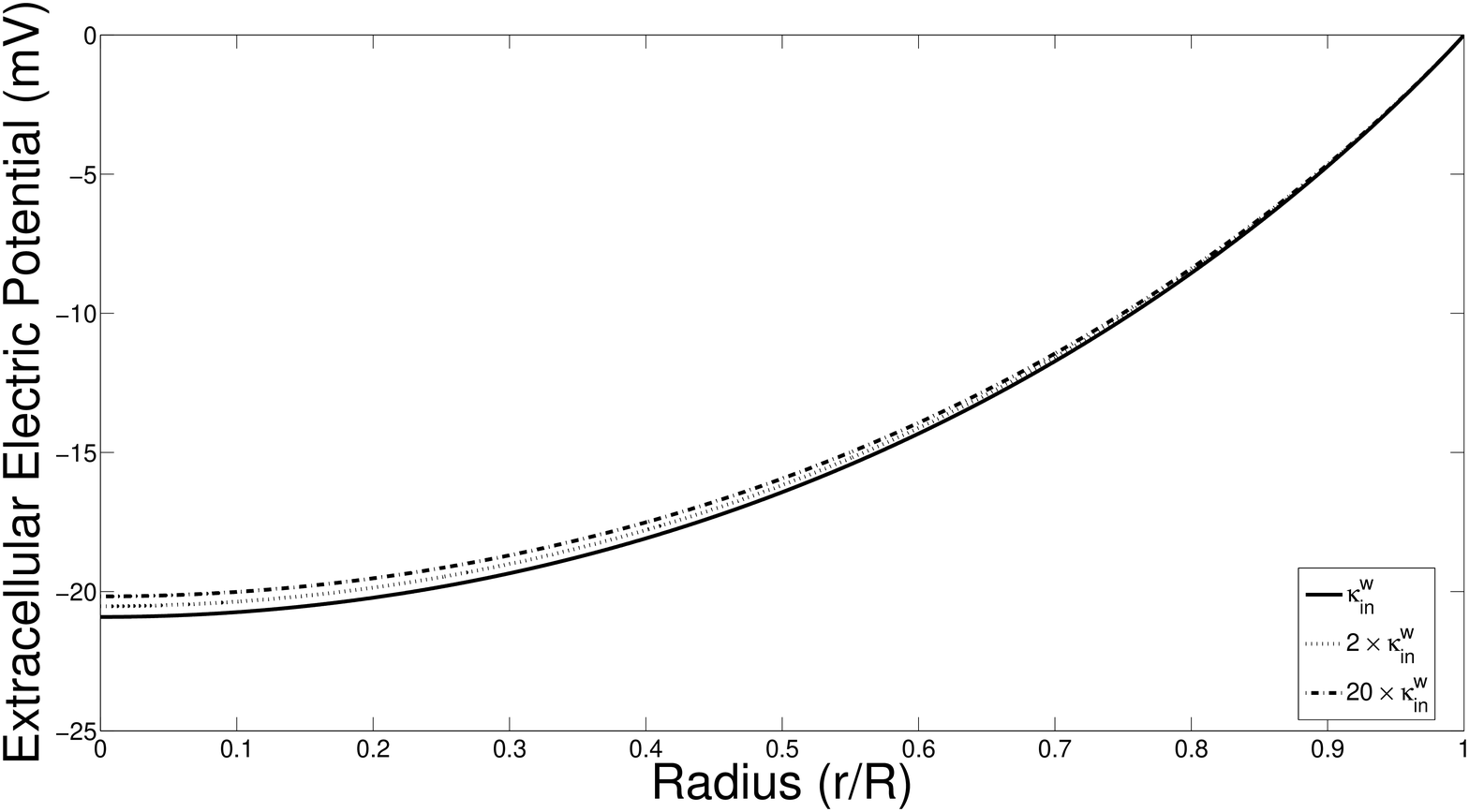}
	\end{minipage}
	\hfill\begin{minipage}[c]{0.45\textwidth}
		\centering
		\includegraphics[width=3.5in,height=2.16in]{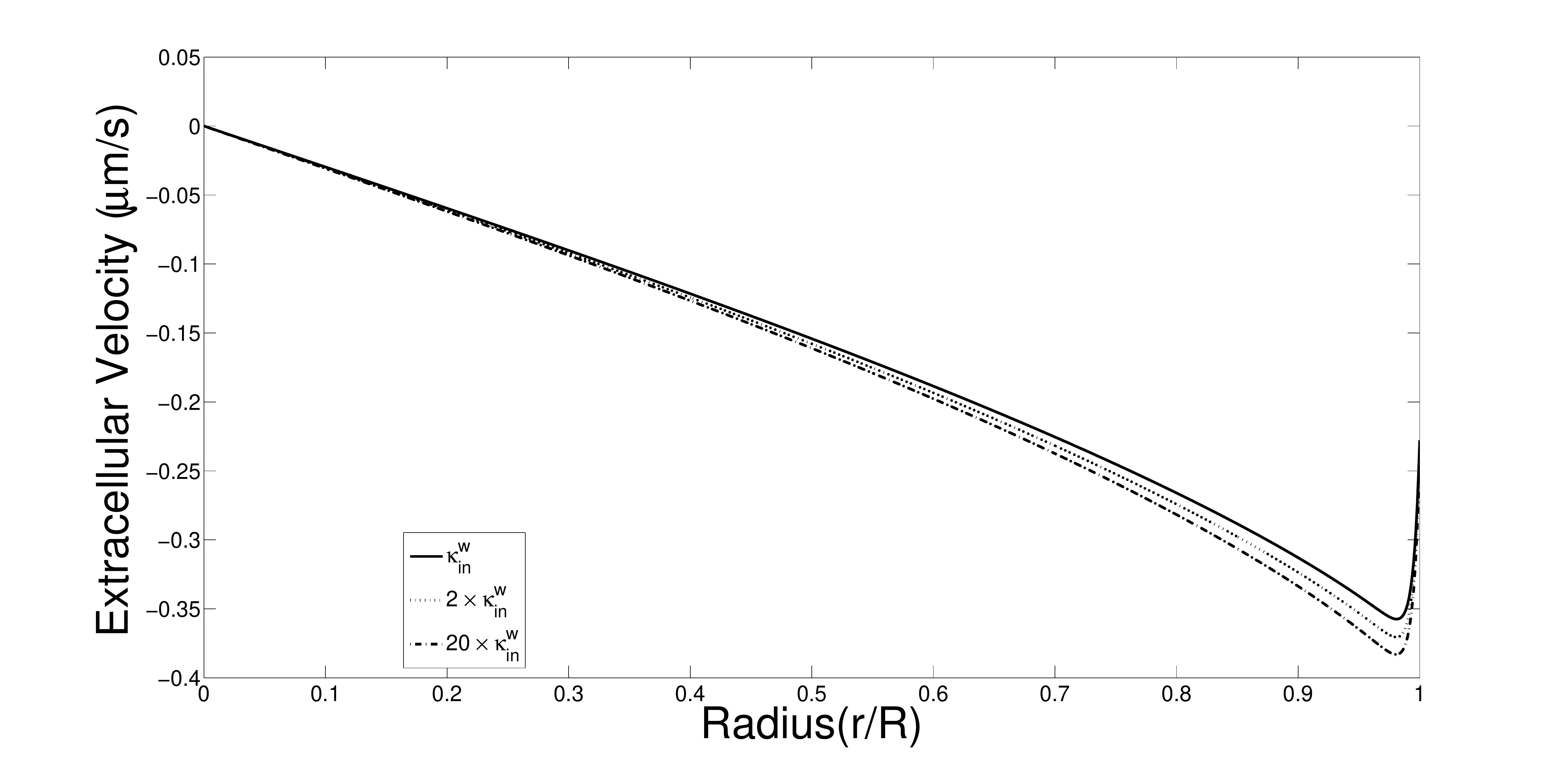}
	\end{minipage}
	\caption{Comparison between different $\kappa_{in}$.}
	\label{figure7}
\end{figure}

\end{document}